\documentclass[useAMS,usenatbib]{mn2e}
\usepackage{amsmath}
\usepackage{amssymb}
\usepackage{graphicx}

\usepackage{graphicx}
\def\cf{{cf.}} 
\def\eg{{e.g.}} 
\def\ie{{i.e.}} 

\def\s{{\rm s}} 
\def\eV{{\rm eV}} 
\def\MeV{{\rm M}\eV} 
\def\erg{{\rm erg}} 
\def\J{{\rm J}} 
\def\G{{\rm G}} 

\def\d{{\rm d}}

\def\dV{{\d^3\!x}}
\def\dS{{\d\bmath{S}}}
\def\dA{{\d^2\!x}}
\def\dC{{\bmath{\d\ell}}}

\def\B{{\bmath{\rm B}}}
\def\E{{\bmath{\rm E}}}
\def\A{{\bmath{\rm A}}}
\def\J{{\bmath{\rm J}}}
\def\blambda{{\bmath{\lambda}}}

\def\hn{{\bmath{\hat{\rm n}}}}
\def\hx{{\bmath{\hat{\rm x}}}}

\def\hz{{\bmath{\hat{\rm z}}}}

\def\curl{{\bmath{\nabla}\times}}
\def\div{{\bmath{\nabla}\cdot}}
\def\grad{{\bmath{\nabla}}}

\def\bY{\bmath{\rm Y}}
\def\bPsi{\bmath{\rm \Psi}}
\def\bPhi{\bmath{\rm \Phi}}

\def\er{\bmath{\hat{\rm e}}_r}
\def\eth{\bmath{\hat{\rm e}}_\theta}
\def\eph{\bmath{\hat{\rm e}}_\phi}

\def\bnabla{\bmath{\nabla}}

\def\cB{{\cal B}}
\def\cA{{\cal A}}

\def\ds{\displaystyle}

\title[Relaxation of Fossil Fields]{Magnetic Helicity and the Relaxation of Fossil Fields}
\author[Avery E. Broderick and Ramesh Narayan]{Avery E. Broderick$^{1}$\thanks{E-mail: abroderick@cfa.harvard.edu (AEB); rnarayan@cfa.harvard.edu (RN)} and Ramesh Narayan$^{1}$\footnotemark[1]\\$^{1}$ Institute for Theory and Computation, Harvard-Smithsonian Center for Astrophysics, MS 51, 60 Garden Street, Cambridge, MA 02138, USA}

\begin{document}

\maketitle

\begin{abstract}
In the absence of an active dynamo, purely poloidal magnetic field
configurations are unstable to large-scale dynamical perturbations,
and decay via reconnection on an Alfv\'enic timescale.  Nevertheless,
a number of classes of dynamo-free stars do exhibit significant,
long-lived, surface magnetic fields.  Numerical simulations suggest
that the large-scale poloidal field in these systems is stabilized by
a toroidal component of the field in the stellar interior.  Using the
principle of conservation of total helicity, we develop a
variational principle for computing the structure of the magnetic
field inside a conducting sphere surrounded by an insulating vacuum.
We show that, for a fixed total helicity, the minimum energy
state corresponds to a force-free configuration.  We find a simple
class of axisymmetric solutions, parametrized by angular and radial
quantum numbers.  However, these solutions have a discontinuity in the
toroidal magnetic field at the stellar surface which will exert a
toroidal stress on the surface of the star.  We then describe two
other classes of solutions, the standard spheromak solutions and ones
with fixed surface magnetic fields, the latter being relevant for
neutron stars with rigid crusts.  We discuss the implications of our
results for the structure of neutron star magnetic fields, the decay
of fields, and the origin of variability and outbursts in magnetars.
\end{abstract}


\section{Introduction}
The importance of magnetic helicity has long been appreciated in the
context of force-free plasma dynamics \citep{Wolt:58,Wolt:62}.  This
is due primarily to the fact that helicity has been experimentally
shown to be conserved during magnetic reconnection
\citep{Tayl:74,Ji-Prag-Sarf:95,Heid-Dang:00,Hsu-Bell:02}, even when a
substantial fraction of the magnetic energy has been dissipated.

In the context of stars, numerical magnetohydrodynamics simulations of
initially random fields in a conducting sphere reveal that magnetic
helicity plays a crucial role in determining the final
magnetic configuration \citep{Brai-Spru:04}.  In particular, after a
period of violent reconnection, which occurs on the Alfv\'en crossing
timescale, a stable, mostly dipolar field develops, the strength of
which depends solely upon the initial helicity.  In the absence of
initial helicity, the final magnetic field vanishes.  
Relevant systems include magnetic A stars, white dwarfs and neutron
stars.  In the present paper we are interested primarily in neutron
stars.

The simplicity of the final solution found by \citet{Brai-Spru:04} and
others \citep[e.g.,][]{Yosh-Yosh-Erig:06} suggests that a simple
analytical model can be developed.  We present here a formalism
that might be relevant for understanding the nature of the final
equilibrium.  Our approach is based on a variational approach that
encodes the principle of conservation of total helicity, as
proposed by \citealt{Tayl:74}.  We do not address the dynamical path that
a system will follow starting from its initial state, but simply
attempt to construct the final state of the system.
Thus, we present a set of magnetic field solutions as a function of
the magnetic helicity.  We obtain these solutions via a variational
energy principle in which the integrated helicity is held fixed.

In \S\ref{HMEatFFC} we state the variational principle and derive the
governing field equations.  \S\ref{AS}, which is the heart of this
paper, then solves the equations under different conditions, assuming
axisymmetry, and discusses the long term evolution of the resulting
magnetic field.  Explicit forms for some illustrative solutions are
given in \S\ref{EEftMF}.  \S\ref{A} discusses the implications of the
results and considers possible applications to neutron stars.  A
number of technical details are given in the Appendices,
including a comparison of our approach, which is based on conservation
of total helicity, and the approach of citep{Fiel:86}, who considers a
conserved microscopic helicity associated with individual magnetic
flux tubes.

\section{Helicity, Minimum Energy, and the Force-Free Condition} \label{HMEatFFC}
The magnetic helicity density is defined by $h\equiv\A\cdot\B$, where
$\A$ is the vector potential and $\B\equiv\curl\A$ is the magnetic
field.  Generally, $h$ is gauge dependent.  Nevertheless, the volume
integrated helicity can be made gauge independent by a suitable choice
of integration.  That is,
\begin{equation}
H \equiv \int_V \frac{1}{8\pi}\A\cdot\B \,\dV \,,
\label{Heq}
\end{equation}
is gauge invariant if $\B\cdot\dS = 0$ everywhere on the boundary of
$V$.  While during reconnection the local helicity density is not
conserved, to a good approximation the total helicity is
\citep{Tayl:74}.  This experimental fact suggests that the
integrated quantity $H$ plays an important role in determining the
final magnetic field configuration, and is the foundation of our
analysis.  In contrast to the helicity, the magnetic energy,
\begin{equation}
U \equiv \int_V \frac{1}{8\pi}\B\cdot\B\,\dV \,,
\label{Ueq}
\end{equation}
is not conserved during reconnection.  Indeed, reconnection can
convert a substantial fraction of the magnetic energy into thermal
energy of the plasma.  

In the absence of a dynamo, the equilibrium state of the field will be
associated with its minimum energy state.  That is, reconnection will
proceed until the maximum amount of magnetic energy is converted into
heat.  If there are no constraints on the system, the entirety of the
magnetic energy will be converted.  However, even under randomly
chosen initial conditions, a non-zero helicity will initially be
present, and subsequently must be conserved.  Thus, there is a limit
to how much magnetic energy can be converted into heat, and the final
configuration must consist of a non-vanishing magnetic field.  This
situation naturally lends itself to a variational approach, which we
explore in the following subsections.

In our analysis we seek to minimize the total magnetic energy
(\ref{Ueq}) subject to a fixed value of the total helicity
(\ref{Heq}).  However, before doing this we might ask: Is it correct
to minimize only the magnetic energy and to exclude contributions to
the energy from the underlying fluid?  The answer is yes.  In the
context of a constrained total helicity, the equations governing
the fluid equilibrium, up to a non-radial surface stress (which we
will assume is balanced by a rigid crust), are completely separable
from those governing the magnetic field structure.  (The reader
is referred to Appendix \ref{RtOVP} for a discussion of this question
and a comparison of our total helicity conservation approach and that
of Field (1986) who considered helicity conservation on individual
field loops).  That is, subject to the constraint of fixed helicity,
we might have considered minimizing
\begin{equation}
\begin{aligned}
U_{\rm tot}
&=
\int_V \rho \left[\epsilon + \frac{1}{2}\Phi\right] \,\dV
+
\int_V \frac{1}{8\pi} \B\cdot\B \,\dV\\
&\equiv
U_f(\rho,V) + U(\B,V)\,,
\end{aligned}
\label{eq:UfB}
\end{equation}
where $\epsilon$ and $\Phi$ are the specific internal energy and
self-gravitational potential of the fluid, respectively, and we assume
for simplicity that the star is barotropic (although this is not
necessary).  Since the helicity constraint depends upon {\em only} the
magnetic field, the magnetic and hydrodynamic contributions to the
energy are distinct, and thus, for a fixed $V$, may be minimized
independently.  As a consequence, the equilibrium magnetic field
configuration will not explicitly depend upon the hydrostatic
structure of the star.  Furthermore, such a state will correspond to a
global minimum of the combined energy, $U_{\rm tot}$.  Perturbations
around such a state will necessarily increase $U_f$ and $U$, otherwise
these would not be independent minima, and thus will increase $U_{\rm
tot}$.  Here and henceforth, when we refer to ``helicity'' we
mean the total helicity defined in equation (\ref{Heq}), as distinct
from the local helicity density $h=\A\cdot\B$ or the helicity
associated with a single magnetic loop (Appendix \ref{RtOVP}).

A somewhat subtle point to note is that the fluid and magnetic field
structure are implicitly coupled by the definition of $V$, the volume
of the conducting region (the star).  That $V$ is finite is critical
to the existence of a stable magnetic field configuration.  Indeed, if
$V$ is infinite there can exist no equilibrium magnetic field
configuration with non-vanishing $\B$.  This is a result of the fact
that $U$ is proportional to $B^2$, and hence the field will naturally
expand to fill all available space and therefore have vanishing strength.
In the case of a star, equilibrium is a consequence of the star's
self-gravity.  This enters into the determination of the magnetic
field by constraining the size of the conducting region, $V$, which
anchors the magnetic field.  Generally, $U_{\rm tot}$ must be
minimized with respect to $V$ as well as $\rho$ and $\B$.  However,
since both $U$ and $U_f$ are functions of $V$, the stellar radius,
$R$, implicitly couples the structure of the magnetic field and the
fluid.  We will discuss how this may be done explicitly in \S\ref{EaH}.

\citet{Wolt:58} demonstrated that, in the context of uniform media,
extremizing the magnetic energy subject to a fixed total helicity
results in the force-free equation,
\begin{equation}
\curl\B=\alpha\B\,,
\label{eq:umec}
\end{equation}
where $\alpha$ is a Lagrange multiplier (Appendix \ref{FFFiUM}).  Such
a field configuration is known as a Woltjer state.  The situation is
more complicated if there are constraints upon the currents that can
be driven in the conducting material.  The simplest case, and the one
of particular interest here, is when the system under consideration
can be separated into two distinct regions: an inner region of
infinite conductivity (type-I region) which corresponds to the
interior of the star, and an outer region of vanishing conductivity
(type-II region) which corresponds to the vacuum exterior of the star.

Given two such regions, the standard variational principle for uniform
media may be extended by the inclusion of Lagrange multipliers
$\alpha$ and $\blambda(\bmath{x})$.  Thus, we seek to extremize the
quantity
\begin{equation}
S = \int_V \frac{1}{8\pi} \left( \B\cdot\B - \alpha\A\cdot\B +
\frac{8\pi}{c} f \blambda\cdot\J \right) \dV\,,
\end{equation}
where the current $\J$ is related to the magnetic field in the normal
way, $\J\equiv c\curl\B/4\pi$, and $f(\bmath{x})$ vanishes in regions
of infinite conductivity (type-I region) and is unity otherwise
(type-II region).  Unlike $\alpha$, which corresponds to fixing $H$
and is a scalar constant, the $\blambda$ are vectors and functions of
position; the difference arises because of the global versus local
nature of their respective constraints.

Varying $S$ with respect to $\A$, holding $\delta\A$ fixed on some
outer boundary ({\em not} on the boundaries between type-I and type-II
regions) gives
\begin{equation}
\curl\B = \alpha\B - \curl\curl f\blambda\,,
\label{eq:nmec}
\end{equation}
and varying with respect to $\lambda$ trivially gives
\begin{equation}
f \curl\B = 0\,.
\label{eq:lam}
\end{equation}
Within type-I ($f=0$) regions, equations (\ref{eq:nmec}) \&
(\ref{eq:lam}) reduce identically to the Woltjer force-free condition
(\ref{eq:umec}).  In contrast, within type-II ($f=1$) regions, the
additional terms are required to make equations (\ref{eq:nmec}) and
(\ref{eq:lam}) consistent with each other.  In particular, inserting
the latter into the former gives
\begin{equation}
\alpha\B = \curl\curl\blambda\,.
\end{equation}
This implies that
\begin{equation}
\curl\blambda = \alpha\A + \alpha\grad\Lambda
\end{equation}
for some $\Lambda$, the explicit form of which depends upon the gauge
choices for $\A$ and $\blambda$.  A natural gauge for the former is
the so-called Force-Free Gauge in which $\B=\alpha\A$ (see Appendix
\ref{FFFiUM}).  This happens also to be a Lorentz gauge (\ie, since we
are concerned with stationary solutions, $\div\A=0$).

As discussed in detail in Appendix \ref{BC}, the discontinuous nature
of $f$ at the boundaries between type-I and type-II regions implies
the standard set of boundary conditions:
\begin{equation}
\begin{aligned}
\hn\cdot\left(\B_{\rm I} - \B_{\rm II}\right) &= 0\,,\\
\hn\times\left(\B_{\rm I} - \B_{\rm II}\right) &= \frac{4\pi}{c}
\bmath{K}\,.
\end{aligned}
\label{eq:bcs}
\end{equation}
Generally, for each choice of $\alpha$ and $\Lambda$, a family of
solutions which are local minima of magnetic energy subject to the
helicity constraint may be produced.  The global minimum can then be
found by inspection.

In the case of a uniform medium it is possible to show that the
magnetic energy and helicity are related by
\begin{equation}
U=\alpha H.
\label{UequalH}
\end{equation}  
In contrast, for non-uniform media no such simple relation exists.
This is because the force-free gauge is not generally a Lorentz gauge
in the type-II regions.  However, we can parametrize the departure
from the uniform-medium expression in terms of the gauge
transformation that relates a Lorentz gauge in the type-II regions to
that determined by the continuity of $\A$ at the boundaries.

In particular, let $\varphi$ be the field potential of the magnetic field
in type-II regions (\ie, $\B=\grad\varphi$) and $\Lambda_{\rm L}$ be a
gauge function which relates $\A$ to the Lorentz gauge in type-II
regions (\ie, $\A=\A_{\rm L}+\grad\Lambda_{\rm L}$ where
$\div\A_{\rm L} = 0$).  Then, as shown in Appendix \ref{TAiNUM}, $U$
and $H$ are related by
\begin{equation}
U -\alpha H
=
\oint_{\partial({\rm I-II})} \frac{\alpha}{8\pi}
\left( \Lambda_{\rm L} \grad\varphi - \varphi \grad\Lambda_{\rm L} \right)\cdot\dS\,,
\label{eq:SofLp}
\end{equation}
which may be identified as the action.

\section{Axisymmetric Solution} \label{AS}
The discussion in the preceeding section was fairly general.  However,
in the context of stellar magnetic field evolution, the case of
interest is the magnetic field of a spherical conducting region
surrounded by an insulating vacuum.  Therefore, henceforth we restrict
our attention to the case in which
\begin{equation}
f(\bmath{x}) = \left\{
\begin{aligned}
&0 && \text{if } |\bmath{x}| \le R \,,\\
&1 && \text{otherwise}\,,
\end{aligned}
\right.
\end{equation}
where $R$ is the stellar radius.  We explicitly construct the solution
using vector spherical harmonics (Appendix \ref{VSH}):
\begin{equation}
\bY_{lm} \equiv \er Y_{lm}\,,\,\,
\bPsi_{lm} \equiv r \bnabla Y_{lm}\,,\,\,
\bPhi_{lm} \equiv \bmath{r}\times\bnabla Y_{lm}\,.
\end{equation}
We also make the further simplification of assuming that the final
solution is axisymmetric.

Assuming axisymmetry simplifies the computation of the force-free
solutions considerably, since the magnetic field can be written as
\begin{equation}
\B = \curl\left(A^\phi\eph\right) + B^\phi\eph\,.
\label{eq:Bdef}
\end{equation}
The general procedure consists of solving for $A^\phi$ and $B^\phi$
for a force-free magnetic field in the stellar interior and matching
this to a vacuum field in the exterior.

\subsection{General Solution} \label{sec:gs}
\subsubsection{Vacuum Exterior} \label{sec:ve}
In the exterior, the magnetic field is defined by equations
(\ref{eq:Bdef}) and (\ref{eq:lam}):
\begin{equation}
\curl\B
=
\curl\curl\left(A^\phi\eph\right) + \curl\left(B^\phi\eph\right)
=
0\,.
\end{equation}
We expand $A^\phi\eph$ and $B^\phi\eph$ in terms of the vector
spherical harmonics,
\begin{equation}
A^\phi\eph
=
\sum_l \cA_l \bPhi_l\,,
\quad
B^\phi\eph
=
\sum_l \cB_l \bPhi_l\,,
\end{equation}
where we have dropped the azimuthal quantum number $m$ because
of axisymmetry ($m=0$).  We then obtain
\begin{equation}
\Delta_l \cA_l = 0\,,\quad \cB_l = 0\,,
\label{eq:vac_as}
\end{equation}
where $\Delta_l$ is the 3-dimensional Laplacian associated with
the meridional harmonic $l$, i.e.,
\begin{equation}
\Delta_l = \frac{1}{r^2} \partial_r r^2 \partial_r - \frac{l(l+1)}{r^2}\,.
\label{eq:Deltal}
\end{equation}
This has the general solution
\begin{equation}
\cA_l = \frac{a_l}{r^{l+1}} + b_l r^l\,.
\end{equation}
Regularity at infinity requires $b_l=0$, and thus the exterior field
is given by
\begin{equation}
\B = \sum_l \left[
- l(l+1) \frac{a_l}{r^{l+2}} \bY_l
+ l \frac{a_l}{r^{l+2}} \bPsi_l
\right]\,.
\label{eq:vacB}
\end{equation}
Note that the exterior $\B$ is purely poloidal.

\subsubsection{Stellar Interior} \label{sec:si}
In the stellar interior it is possible to make use of the force-free
gauge ($\B=\alpha\A$), and rewrite $\B$ in terms of  $B^\phi$ alone:
\begin{equation}
\B = \alpha^{-1} \curl\left(B^\phi\eph\right) + B^\phi\eph\,.
\end{equation}
The vector field $B^\phi\eph$ is again expanded in vector spherical
harmonics, with the result that the force-free equation is
\begin{equation}
\curl\curl \sum_l \cB_l \bPhi_l
=
\alpha \sum_l \cB_l \bPhi_l\,.
\end{equation}
Upon making use of the properties of the vector spherical harmonics,
this gives
\begin{equation}
\Delta_l \cB_l + \alpha^2 \cB_l = 0\,.
\label{eq:ffeq_bs}
\end{equation}

The general solution of equation (\ref{eq:ffeq_bs}) can be written in
terms of spherical Bessel functions:
\begin{equation}
\cB_l = \alpha\left[ c_l n_l(\alpha r) + d_l j_l(\alpha r) \right]\,,
\end{equation}
where $j_l$ and $n_l$ are the spherical Bessel functions of the first
and second kind, respectively.  Regularity at the centre and the
orthogonality of the $n_l$ imply that the $c_l$ must vanish.
Therefore, the general solution for the axisymmetric force-free
magnetic field in the stellar interior is
\begin{multline}
\B = \sum_l \left\{
-\left[\frac{l(l+1)}{r} d_l j_l(\alpha r)\right] \bY_l\right.\\
-\left[\frac{1}{r}\partial_r r d_l j_l(\alpha r)\right] \bPsi_l
\left.+ \alpha d_l j_l(\alpha r) \bPhi_l
\right\}\,.
\label{eq:ffB}
\end{multline}
In general, the interior field has both poloidal and toroidal
components.

\subsubsection{Matching Condition} \label{sec:mc}

Equations (\ref{eq:vacB}) and (\ref{eq:ffB}) give the general form of
the magnetic field in the vacuum exterior and the conducting stellar
interior.  Each is expressed in terms of a set of coefficients, $a_l$
and $d_l$, respectively.  The two solutions must be matched across the
stellar surface, $r=R$, using the standard boundary conditions
(equation \ref{eq:bcs}).  This gives the following relation between
the $a_l$ and $d_l$:
\begin{equation}
d_l = \frac{a_l}{R^{l+1} j_l(\alpha R)}\,.
\label{eq:match}
\end{equation}

\subsection{Minimum Energy Solutions}\label{sec:minenergy}
The set of field solutions (\ref{eq:vacB}) and (\ref{eq:ffB}) with the
matching condition (\ref{eq:match}) all satisfy the equations obtained
from varying $S$ with respect to $\A$.  However, we have yet to
determine the minimum energy state.  That is, we have replaced the
degrees of freedom associated with the functional form of $\B$ with an
infinite, but countable, set of degrees of freedom associated with the
coefficients $a_l$.  Furthermore, we have yet to choose an $\alpha$.
This can be addressed by considering $S(a_l;\alpha)$ explicitly, \ie,
minimizing the energy with respect to the $a_l$ and $\alpha$, subject
to the constraint of fixed $H$.

Equation (\ref{eq:SofLp}) provides a direct way in which to construct
$S(a_l;\alpha)$ in terms of $\varphi$ and $\Lambda_{\rm L}$, and thus
we will determine functional forms for these now.  By inspection of
equation (\ref{eq:vacB}),
\begin{equation}
\varphi
=
\sum_l \frac{a_l l}{r^{l+1}} Y_{l0}\,.
\end{equation}
Determining $\Lambda_{\rm L}$ is more difficult.  Letting $\A = A^\phi
\eph$ is sufficient to produce the required vacuum field solution, and
also satisfies the Lorentz gauge.  However, $\A$ must be continuous
across the stellar surface.  Thus, $\Lambda_{\rm L}$ is defined by
\begin{multline}
\left. \grad\Lambda_{\rm L} \right|_{r=R}
=
\sum_l \frac{d_l}{\alpha} \left\{
-
\left[\frac{l(l+1)}{R} j_l(\alpha R)\right] \bY_l \right.\\
\left.
+
\left[
\frac{l}{R} j_l(\alpha R) - \alpha j_{l-1}(\alpha R)
\right]\bPsi_l
\right\}\,,
\end{multline}
where $\partial_z z j_l(z) = z j_{l-1}(z) - l j_l(z)$ was used
\citep{Abra-Steg:72}.  Setting $\Lambda_{\rm L} = \sum_l \lambda_{{\rm
L}l}(r) Y_{l0}(\theta,\phi)$ gives
\begin{equation}
\grad\Lambda_{\rm L}
=
\sum_l \left[
\partial_r \lambda_{{\rm L}l} \bY_l
+
\frac{\lambda_{{\rm L}l}}{r} \bPsi_l
\right]\,,
\end{equation}
which, together with the orthogonality of the vector spherical
harmonics, implies
\begin{equation}
\lambda_{{\rm L}l}
=
\frac{d_l}{\alpha} \left[ \left(\frac{r}{R}\right)^{-(l+1)} l j_l(\alpha R)
-
\alpha R j_{l-1}(\alpha R) \right]\,.
\end{equation}

Making use of the above results, and after integrating over the
stellar surface, we obtain
\begin{equation}
S(a_l;\alpha) =
\frac{\alpha}{8\pi} 
\sum_l a_l^2  \frac{l(l+1)}{R^{2 l}}\frac{j_{l-1}(\alpha R)}{j_l(\alpha R)}\,.
\label{eq:action_bs}
\end{equation}
As expected, variations of $S(a_l;\alpha)$ with respect to $\alpha$
gives back $H(a_l;\alpha)$ (\cf~with equation \ref{eq:helicity_ff}).
Variations with respect to $a_l$ gives a new condition,
\begin{equation}
\alpha a_l \frac{j_{l-1}(\alpha R)}{j_l(\alpha R)} = 0\,,
\label{eq:condition}
\end{equation}
which implies that either $a_l=0$ or $j_{l-1}(\alpha R)=0$ for all
$l$.  Since the zeros of the spherical Bessel functions are distinct,
at most one multipole component can be non-vanishing.

Thus, we conclude that a minimum energy configuration consists of a
single multipole $l$ with the value of $\alpha$ selected such that
\begin{equation}
j_{l-1}(\alpha R)=0, \qquad a_l \ne 0.
\label{eq:jl}
\end{equation}  
It is worth noting that the above condition on $\alpha$ corresponds to
the poloidal component of the field being continuous across the
stellar surface.  That is, perhaps unsurprisingly, the minimum energy
is obtained when there is no kink in the poloidal component of the
magnetic field at the surface.  There is, however, generally a kink in
the toroidal component of the field.  

The toroidal kink at the surface will exert a toroidal stress upon the
surface of the star, the direction of which depends upon the sense of
the magnetic field.  In the case of a purely fluid star, the stress
will cause motions that will result in the untwisting of field lines
which cross the surface and thus a reduction in the overall helicity
of the system.
One way to eliminate this effect is to focus on solutions with zero
stress at the surface.  We describe a class of such solutions in
\S\ref{sec:zerotorque}.  Alternatively, if the star has a solid crust,
it can balance a non-vanishing surface stress by rigidly coupling
different parts of the surface.  We discuss the effect of a crust in
\S\ref{sec:crust} below.  For now we continue our discussion of the
minimum energy solutions given by equation (\ref{eq:jl}), ignoring the
stress at the surface.  As it happens, many of the qualitative
features of these solutions carry through to the other situations.

\begin{figure*}
\begin{center}
\begin{tabular}{cc}
\includegraphics[width=0.5\textwidth]{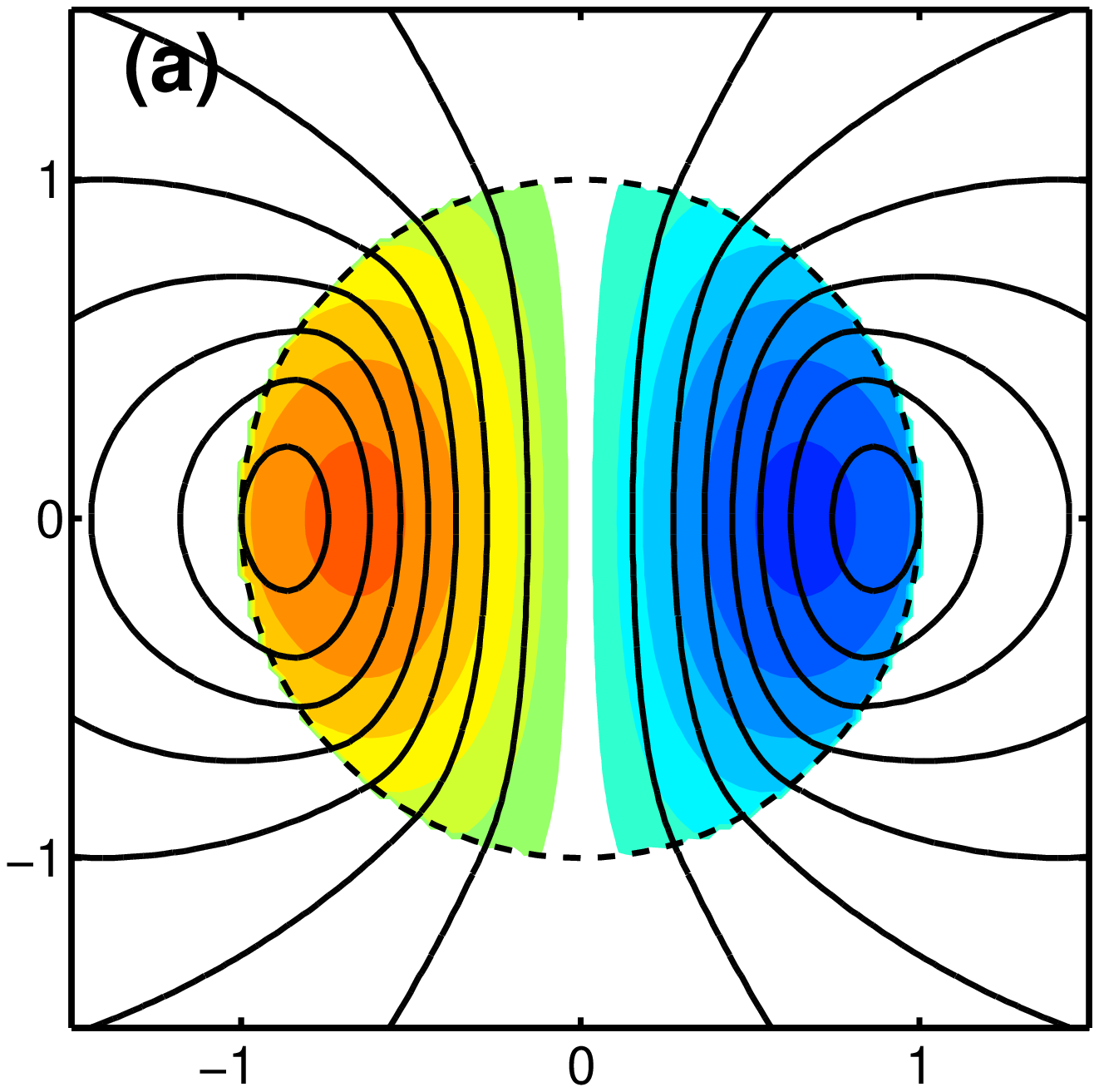}
&
\includegraphics[width=0.5\textwidth]{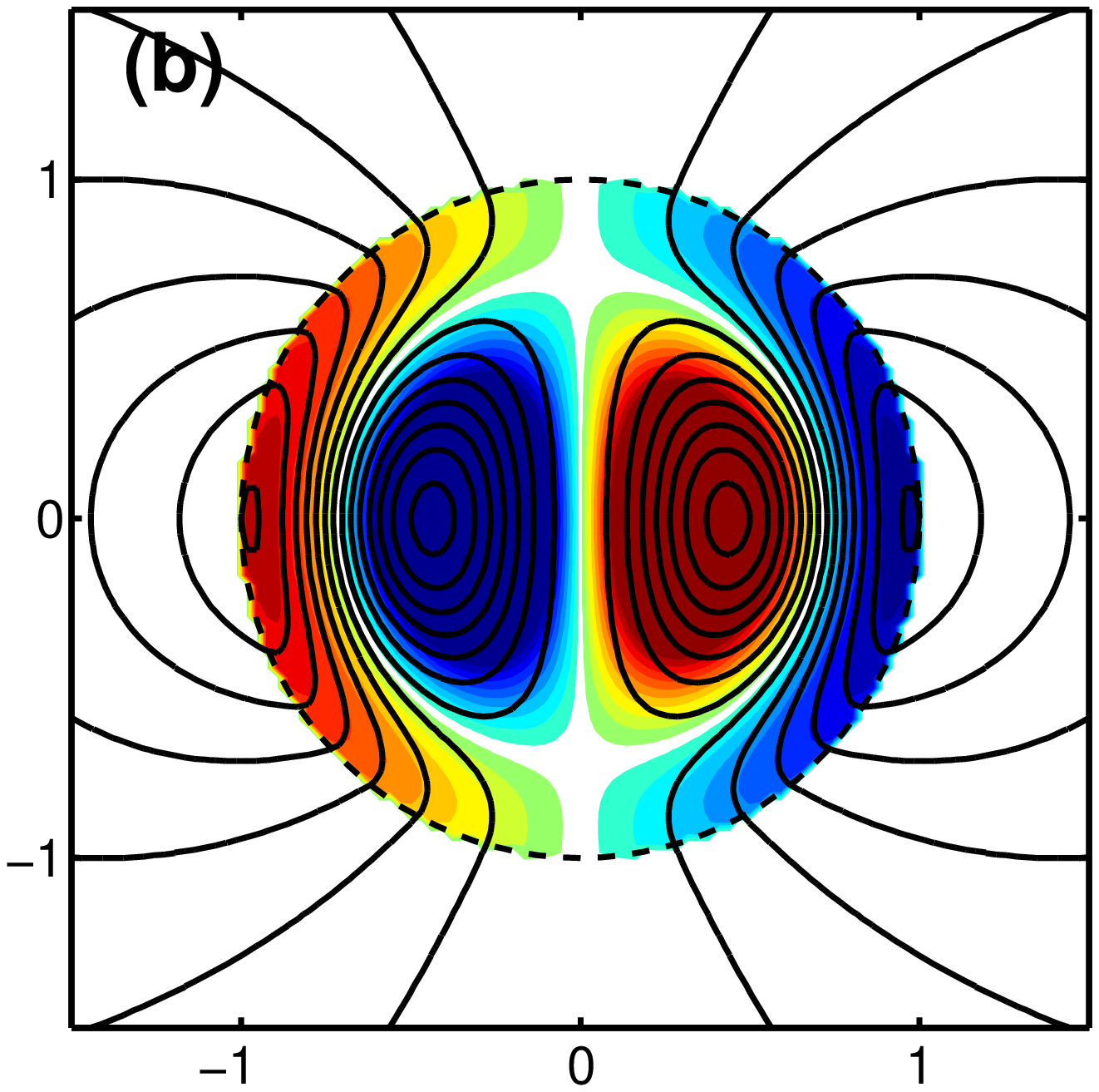}
\\
\includegraphics[width=0.5\textwidth]{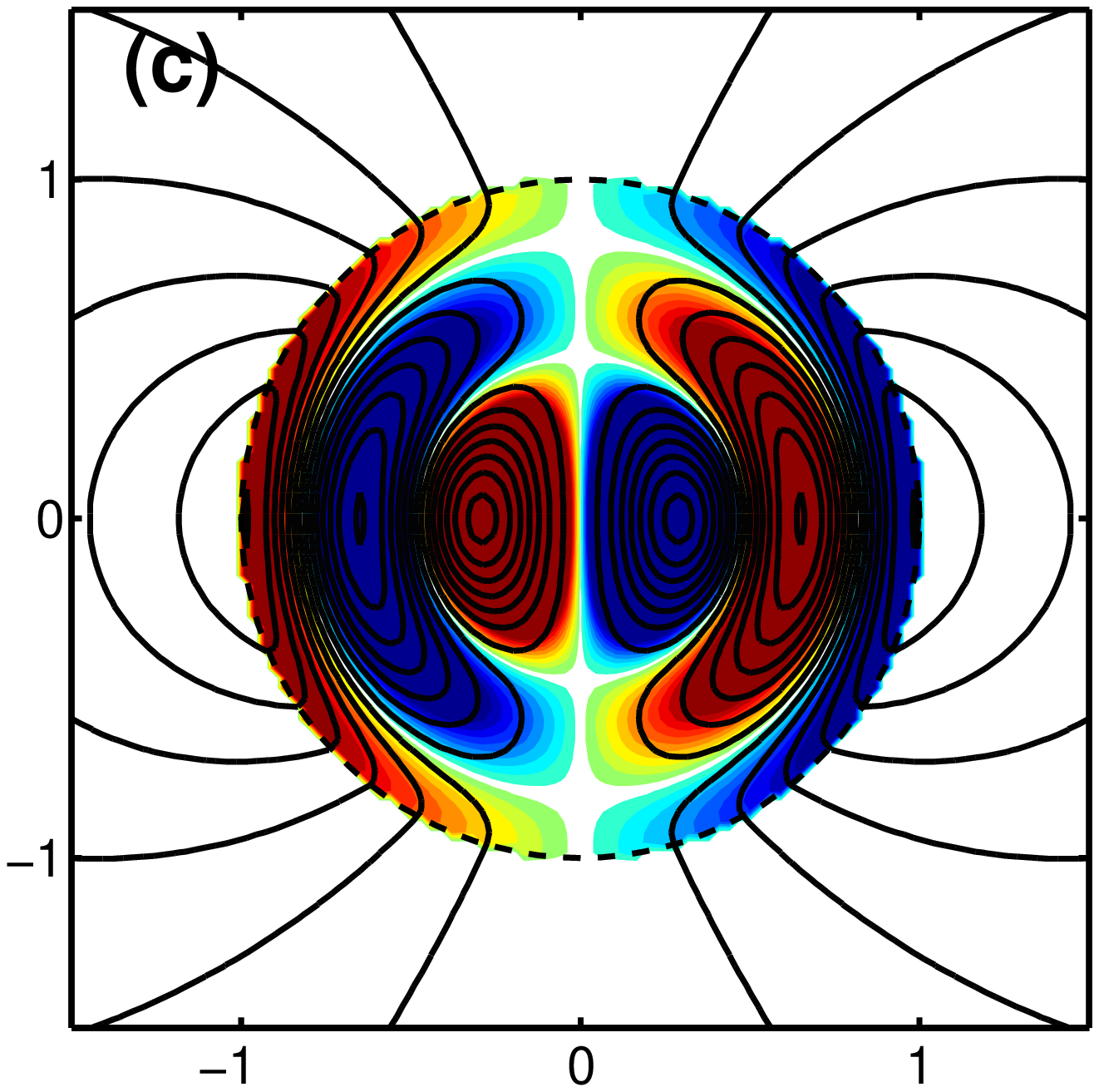}
&
\includegraphics[width=0.5\textwidth]{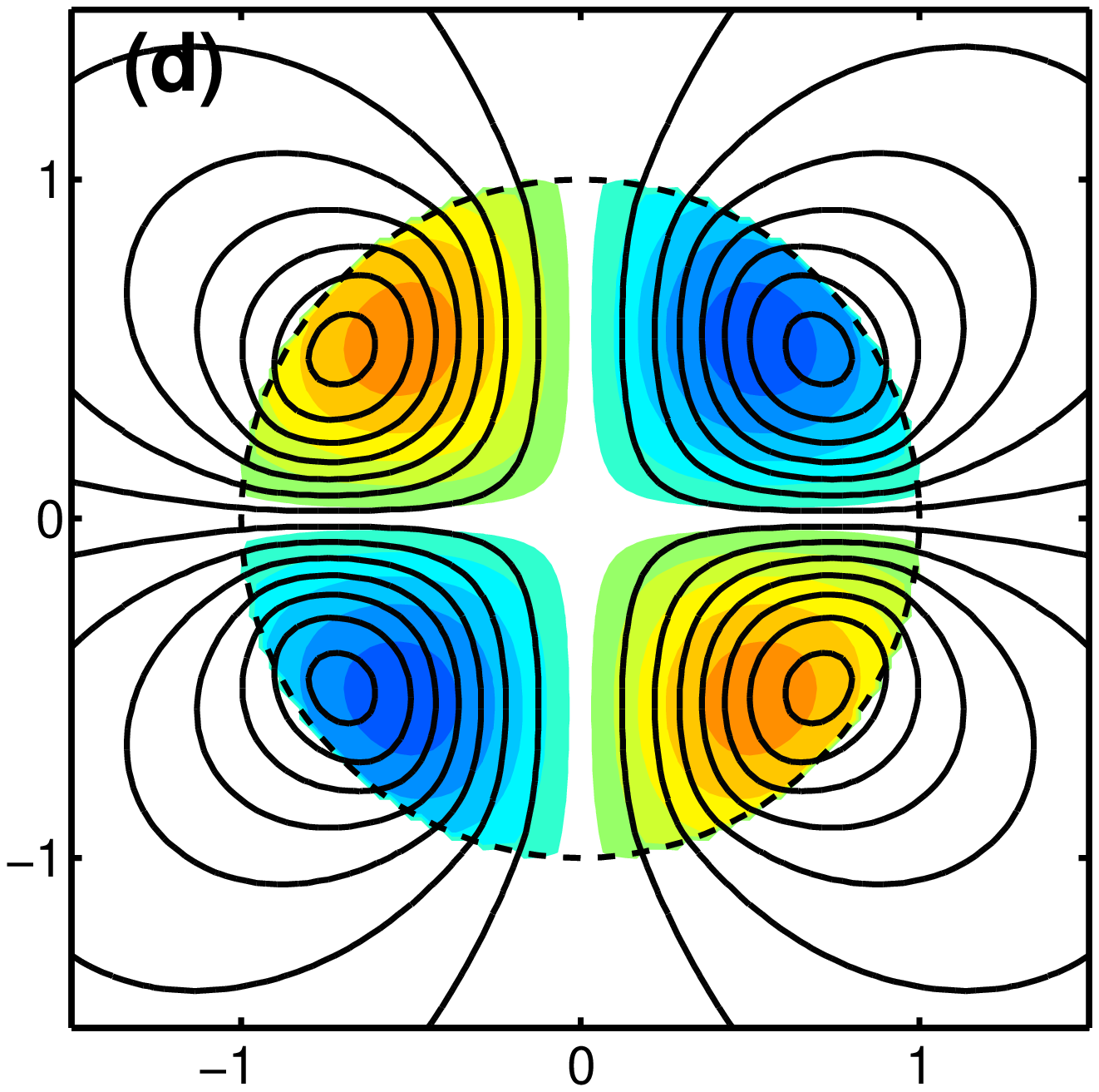}
\end{tabular}
\end{center}
\caption{Meridional slices of illustrative magnetic field geometries
for (a) the global minimum energy configuration, which corresponds to
the first energy minimum of a dipole configuration, (b) \& (c) the
second and third energy minima of a dipole configuration, and (d) the
lowest energy minimum of a quadrupole configuration.  The poloidal and
toroidal field structure are shown by the field lines and filled-color
contours, respectively.  In all plots the field line density is
proportional to the poloidal flux and is normalized such that the
average surface fields are equal.  Similarly, the toroidal color
scheme is uniform among all of the plots.}
\label{fig:bfields}
\end{figure*}

\subsection{Energy and Helicity} \label{EaH}
For the values of $a_l$ and $\alpha$ found in the previous subsection,
$S$ identically vanishes (substitute eq. \ref{eq:condition} in
eq. \ref{eq:action_bs}).  As a consequence, in an energy minimum
state, the magnetic helicity and energy are related by equation
(\ref{UequalH}), exactly as for Woltjer states in uniform media. After
integrating equation (\ref{eq:Utot}) for the magnetic energy, we find
that
\begin{equation}
U = \alpha H = l(l+1) \alpha^2 a_l^2 / 8\pi R^{2l-2}\,.
\end{equation}
For a given fixed helicity $H$ and given $l$ and $\alpha$, this
relation determines the value of $a_l$.

We have seen in the previous subsection that we have an infinite
number of minimum energy solutions.  This is because we are free to
choose any value of the multipole index $l$, and for each $l$, we may
choose any zero of $j_{l-1}$ (see eq. \ref{eq:jl}) to evaluate
$\alpha$.  Every one of these solutions is a local energy minimum with
respect to variations of all quantities (including the fluid
variables, see \S1).  Of these many local minima, one corresponds to
the global energy minimum for the given magnetic helicity.  Since
$U=\alpha H$, the global minimum will have the smallest value of
$\alpha$ among all the solutions, i.e., the smallest root of
$j_{l-1}(z)$ for $l\geq0$.  By inspection it is clear that the global
minimum energy solution has $l=1$ and $\alpha=\pi/R$, which
corresponds to the first zero of $j_0(z)$.  The magnetic field
configuration of this solution corresponds to a dipole and is shown in
Figure \ref{fig:bfields}a.  As mentioned, there are countless higher
energy solutions, each of which corresponds to a local minimum.  These
involve higher multipoles and/or more complex radial structure.  Some
examples are shown in panels b--d of Figure \ref{fig:bfields}, and are
discussed in more detail in the following subsection.

As an aside, we note that these equilibrium configurations describe a
class of solutions, defined by $\alpha R$ and $H$, as a function of
$R$.  Since $\alpha R$ and $H$ are fixed, $U\propto R^{-1}$.
Therefore, if we were free to minimize $U$ with respect to $R$, we
would find that $R\rightarrow\infty$ and $B\rightarrow 0$.  This does
not happen in practice because $R$ is also constrained by the
self-gravity of the star.  That is, we must minimize $U_{\rm tot} =
U_f + U$ (see eq. \ref{eq:UfB}) with respect to $R$, not $U$ alone.
Generally, $U_f$ cannot be a monotonic function of $R$, otherwise no
stable unmagnetized hydrostatic configuration would exist (which is a
necessary a priori assumption).  To illustrate this procedure, let us
assume that $U\ll U_f$, and thus the correction to $R$ associated with
the magnetic pressure is small (which will be shown below).  Then, we
may approximate $U_f$ and $U$ by
\begin{equation}
U_f \simeq U_{f,0} + \frac{1}{2} \frac{\partial^2 U_f}{\partial R^2} \delta R^2
\quad\text{and}\quad
U \simeq U_{0} - U_{0} \frac{\delta R}{R_0}\,,
\end{equation}
where $\delta R$ is the perturbation in the stellar radius as a result
of the presence of the magnetic field.  Thus, variations with respect
to $\delta R$ give
\begin{equation}
\delta R 
\simeq
\frac{U_{0}/R_0}{\partial^2 U_f/\partial R^2}
\sim
\frac{U_{0}}{U_{f,0}} R_0\,.
\end{equation}
Therefore, subdominant fields will correspond to a small perturbation
to $R_0$ while equipartition fields will make corrections of order unity.
However, it is clear that a stable configuration with finite $R$ and
non-vanishing magnetic field does exist, and may be described by the
analysis presented here.

\subsection{Non-Equilibrium Configurations} \label{NEC}
\begin{figure}
\begin{center}
\includegraphics[width=\columnwidth]{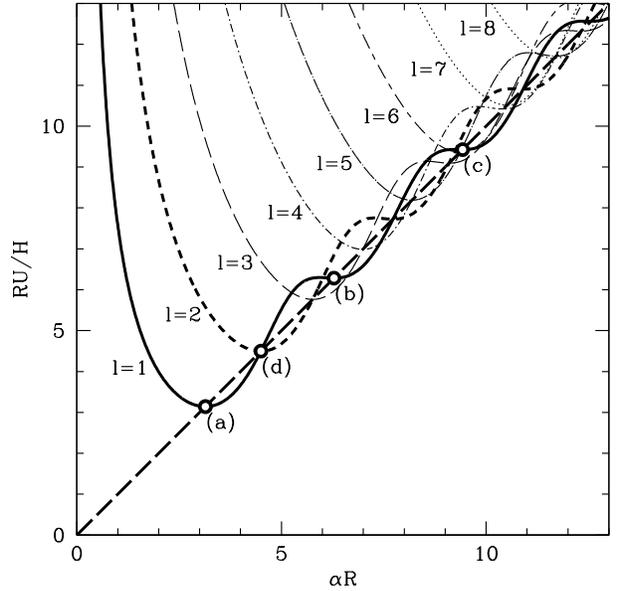}
\end{center}
\caption{The energy per unit helicity as a function of the force-free
constant, $\alpha$, both appropriately scaled by the stellar radius.
The solid, short-dashed and long-dashed lines represent the first
three multipoles, $l=1$, 2, 3, respectively.  The dipole configuration
with $\alpha R = \pi$, labelled (a), corresponds to the global minimum
energy for a given helicity; the associated field geometry is shown in
Figure \ref{fig:bfields}a.  There is a spectrum of higher energy local
minima for each multipole, \eg, (b) and (c) for the dipole, which
correspond to Figures \ref{fig:bfields}b and \ref{fig:bfields}c.  The
point labeled (d) is the minimum energy quadrupole configuration, and
corresponds to Figure \ref{fig:bfields}d.  For reference, the diagonal
line corresponds to $U=\alpha H$.}
\label{fig:UH}
\end{figure}
While we have found a globally optimized magnetic field geometry for a
given magnetic helicity, thus far we have presumed that the star will
naturally seek this minimum energy state via large scale motions and
reconnection events.  However, there is no guarantee that the star
will not become trapped in higher energy, local equilibria, in which the
necessary large scale rearrangements are inhibited.  For this reason we
consider a simple evolution, comprised of a single multipole and
varying $\alpha$, which may occur, for instance, in some forms of
current decay.

For values of $\alpha$ which are not zeros of $j_l(\alpha R)$ the
action does not vanish.  Nonetheless, it is straightforward to compute
$U$ and use $S(\alpha R)$ to relate this to $H(\alpha)$:
\begin{eqnarray}
U &=& \frac{l(l+1)}{8\pi R^{2l-1}} \alpha^2 a_l^2
\bigg[
1
-
\frac{2 l}{\alpha R} \frac{j_{l-1}(\alpha R)}{j_{l}(\alpha R)}
+
\frac{j_{l-1}^2(\alpha R)}{j_{l}^2(\alpha R)}
\bigg],
\label{eq:energy_ff}\\
H
&=&
\frac{U-S}{\alpha} \nonumber \\
&=&
\frac{l(l+1)}{8\pi R^{2l-1}} \alpha a_l^2
\bigg[
1
-
\frac{2 l+1}{\alpha R} \frac{j_{l-1}(\alpha R)}{j_{l}(\alpha R)}
+
\frac{j_{l-1}^2(\alpha R)}{j_{l}^2(\alpha R)}
\bigg]
.
\label{eq:helicity_ff}
\end{eqnarray}
Since equilibria minimize the energy per unit helicity, we show in
Figure \ref{fig:UH} the quantity $R U/H$ as a function of $\alpha R$
for a number of multipoles.  For each multiple there is a global
minimum, with $U/H$ being smallest for the dipole.  However, there are
also a number of local minima corresponding to higher energy states.
The explicit geometry for some of these are shown in Figure 1, and
their implications are discussed in \S\ref{LTFE} and \S\ref{A}.

\subsection{Solutions With no Toroidal Surface Stress}\label{sec:zerotorque}

The solutions discussed so far are true minima of the total energy
subject to a fixed helicity.  Yet, they are not true equilibria of the
system since they have unbalanced stresses at the stellar surface.
This is a paradoxical situation --- how can a minimum energy
configuration not be in equilibrium?  The explanation is that the
motions that the surface stress will induce motions that cause an
unwinding of the toroidal component of the magnetic field and thus a
reduction in the magnetic
helicity of the system.  In effect, the unwinding of the field causes
helicity to flow out of the star to infinity.  Thus, while the
solutions discussed so far are indeed true minimum energy states so
long as helicity is conserved, they are not relevant for a fully fluid
star because such a star does not preserve helicity.

One way to ensure helicity conservation is to make sure that there is
no surface tangential stress, i.e., the toroidal magnetic field in the stellar
interior vanishes as $r\to R$.  By equation (\ref{eq:ffB}), this is
easily arranged by having a field configuration with a single
multipole $l$ and choosing $\alpha$ such that
\begin{equation}
j_l(\alpha R)=0, \qquad a_l \ne 0.
\label{eq:jl2}
\end{equation}
The situation is very similar to that of the minimum energy solutions
discussed earlier, except that we now make use of a root of a
different spherical Bessel function (\cf~eqs. \ref{eq:jl} and
\ref{eq:jl2}).  These are simply the standard
spheromak solutions, though we shall refer to them as zero-torque
solutions to contrast them with the other solutions discussed
throughout this paper.

As before, we have the option of choosing any value of $l$ and any
zero of the corresponding Bessel function.  Thus, we have an infinite
number of possible solutions, of which that corresponding to $l=1$ and
the first zero of $j_1(\alpha R)$ represents the overall lowest energy
state.  All these solutions have vanishing field in the exterior
vacuum, as required by the matching condition (\ref{eq:match}), and
they have no unbalanced tangential stress at the surface of the star.
However, these configurations still require the presence of a rigid
bounding crust.  This is because they do have an unbalanced radial
stress resulting from the discontinuity in the tangential field at the
surface (it is non-zero inside and vanishes outside).  Thus these
configurations are also inappropriate for an entirely fluid star.

\begin{figure*}
\begin{center}
\begin{tabular}{cc}
\includegraphics[width=0.5\textwidth]{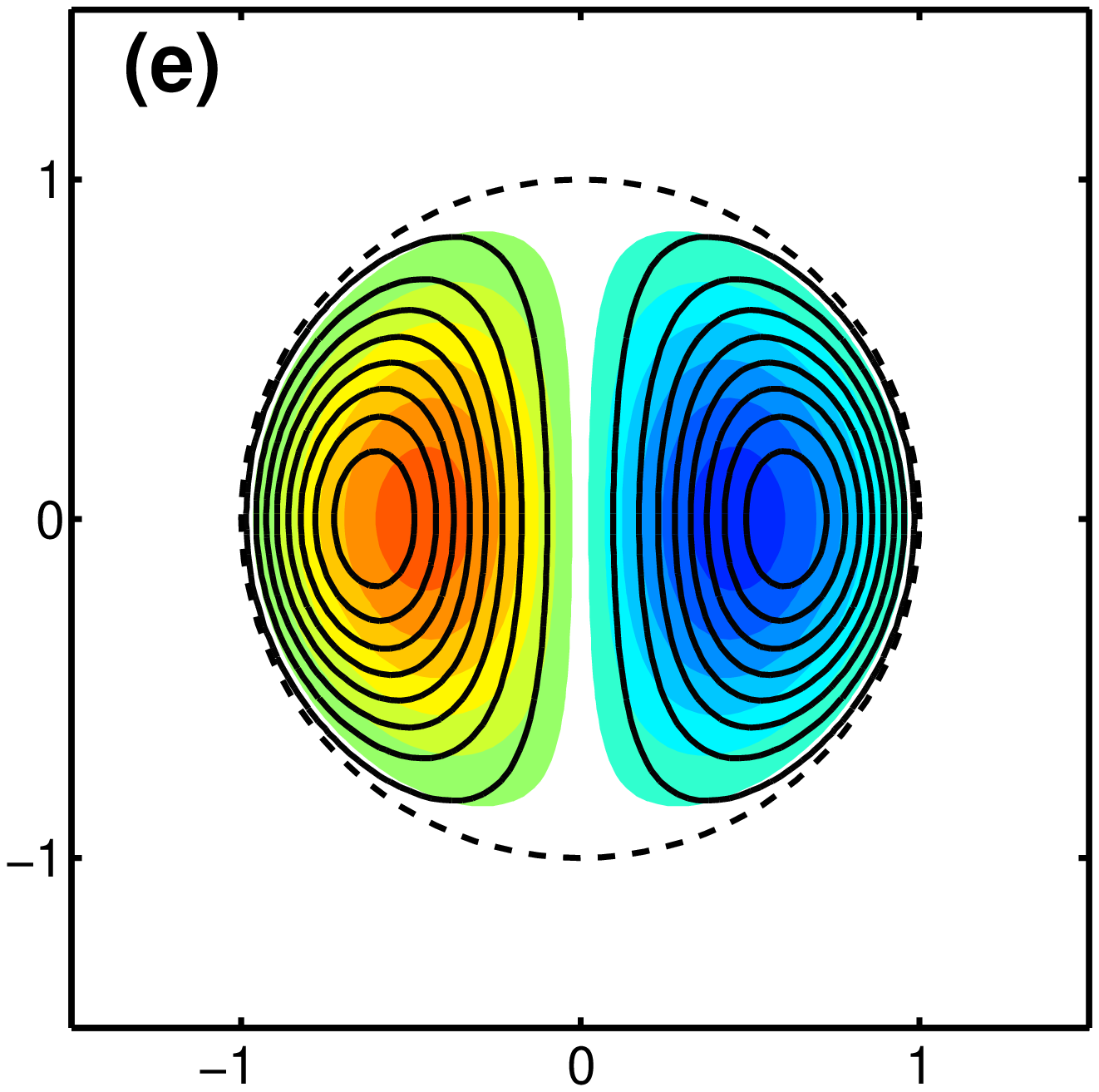}
&
\includegraphics[width=0.5\textwidth]{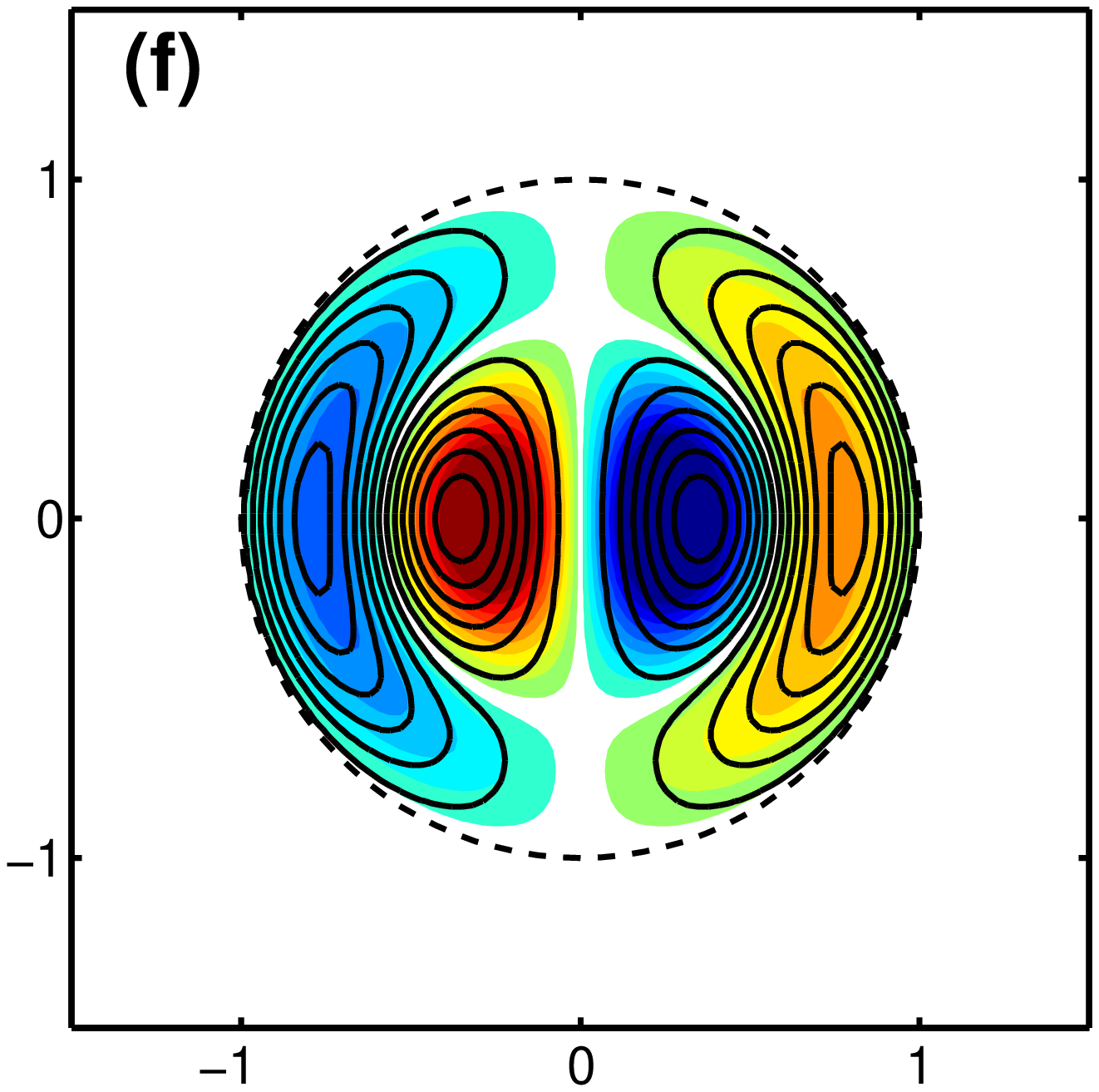}
\\
\includegraphics[width=0.5\textwidth]{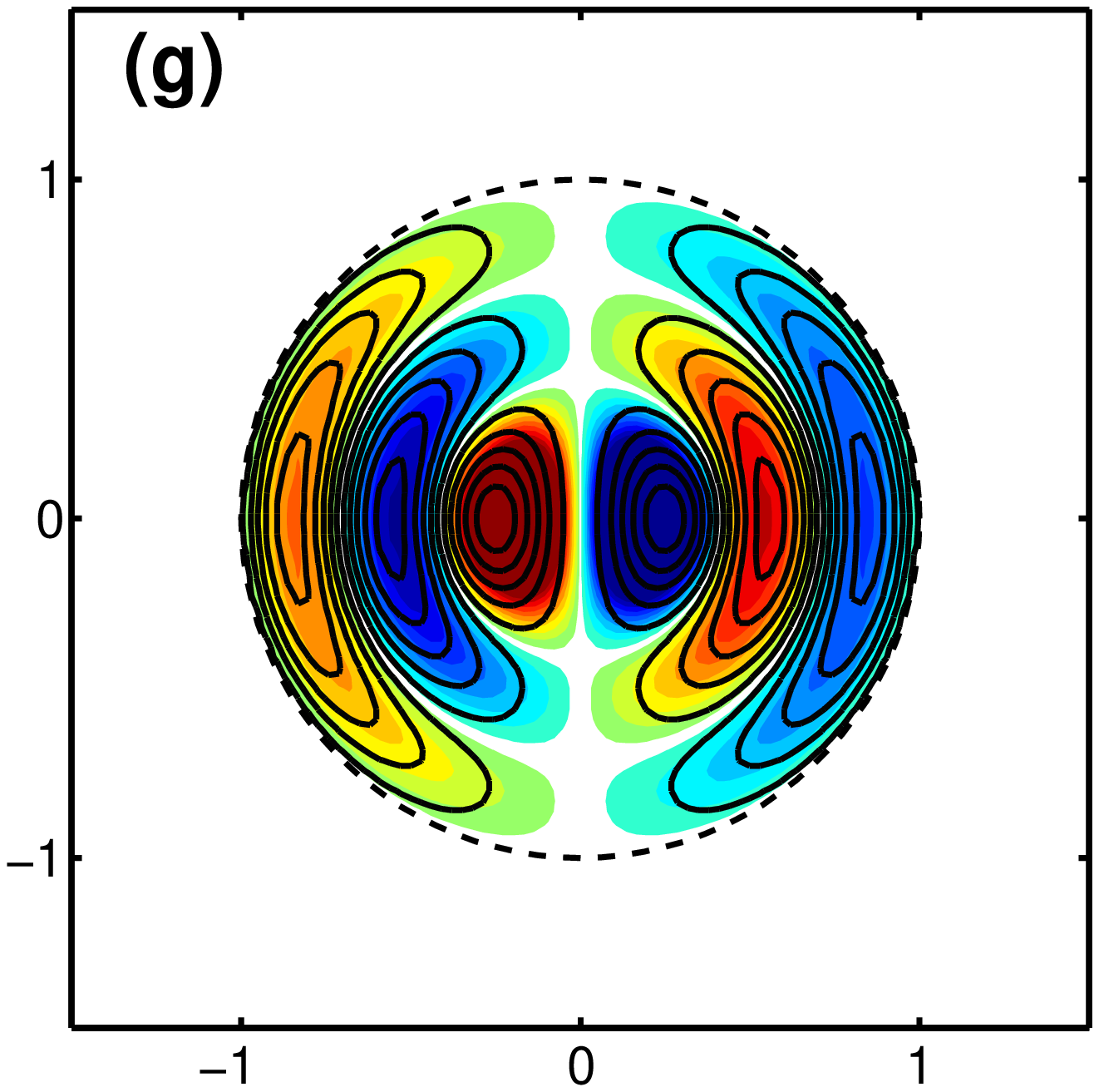}
&
\includegraphics[width=0.5\textwidth]{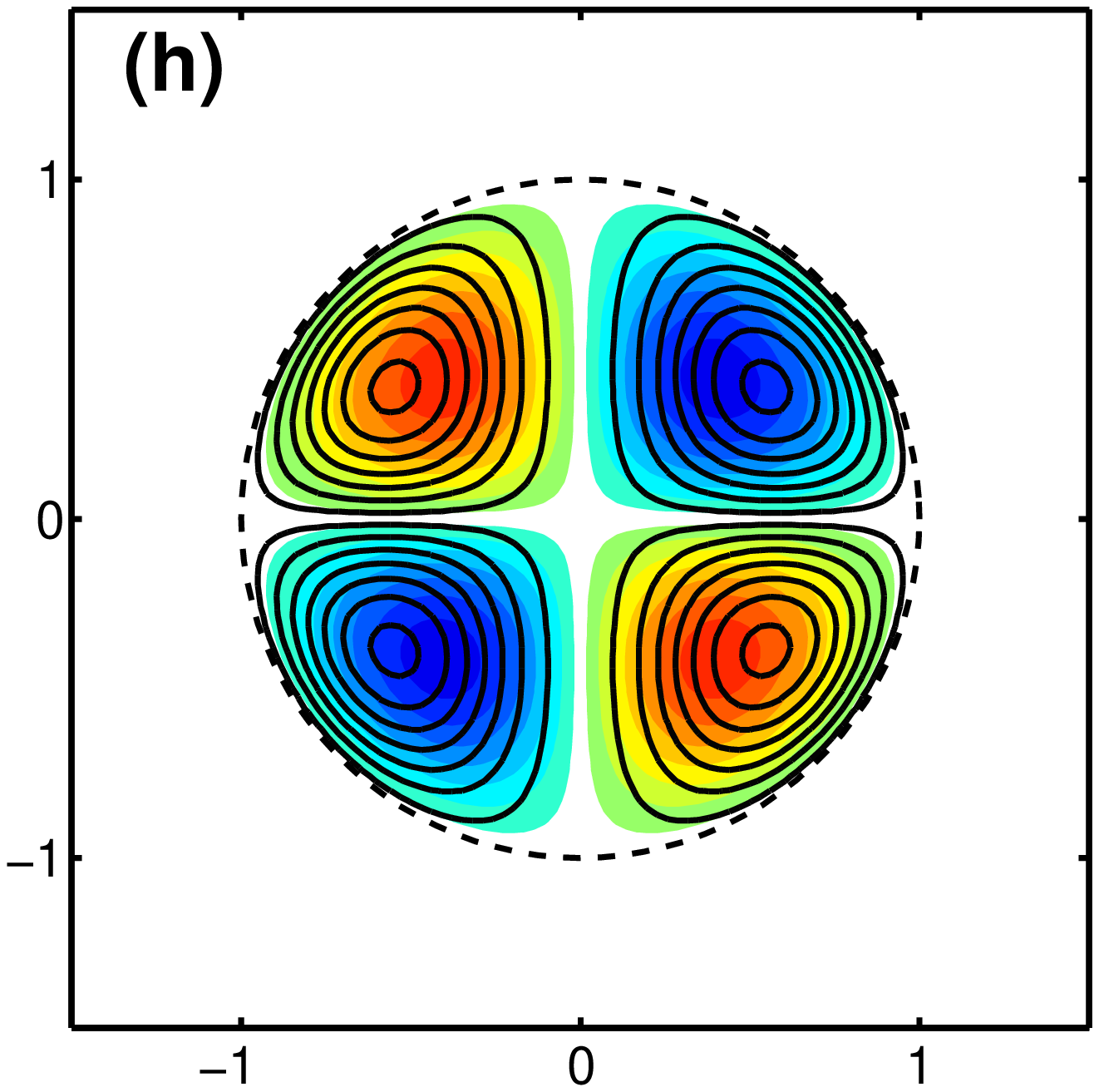}
\end{tabular}
\end{center}
\caption{Meridional slices of illustrative magnetic field geometries
with vanishing surface stresses for (e) the lowest energy such
configuration, corresponding to the first dipole state, (f) \& (g) the
second and third dipole solutions, and (h) the lowest energy
quadrupole configuration.  The poloidal and toroidal field structure
are shown by the field lines and filled-color contours, respectively.
In all plots the field line density is proportional to the poloidal
flux densities and the color scheme is proportional to the toroidal
flux densities.  Both are normalized such that all frames have
identical total helicities.  Note that there is no magnetic field
outside the star.}
\label{fig:zt_bfields}
\end{figure*}

\begin{figure}
\begin{center}
\includegraphics[width=\columnwidth]{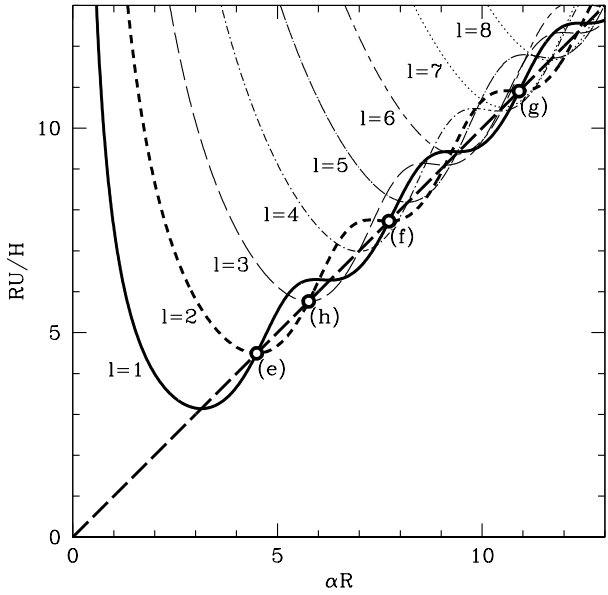}
\end{center}
\caption{The same as Figure \ref{fig:UH}, except the zero-torque
  configurations shown in Figure \ref{fig:zt_bfields} are labeled (e),
  (f), (g) (the first three dipole zero-torque states) and (h) (the
  first quadrupole zero-torque state).  For reference, the diagonal
  line corresponds to $U=\alpha H$.  Note that these solutions are not
  at minima in $U/H$, and thus are likely unstable.}
\label{fig:zt_UH}
\end{figure}

Figure \ref{fig:zt_bfields} shows meridional slices of the magnetic field corresponding
to the first three zero-torque solutions with dipole geometry
($l=1$) and the first quadrupole solution ($l=2$).  These solutions
are very similar to the minimum energy solutions shown in Figure 1.
The main difference is that the magnetic field is contained entirely
inside the star and the field outside the star vanishes.  Figure \ref{fig:zt_UH} is
analogous to Figure 2, except that it highlights the solutions
corresponding to the four panels in Figure \ref{fig:zt_bfields}.  We note two interesting
facts.  First, the new solutions still satisfy $U=\alpha H$; hence,
for a given $H$, the value of $\alpha$ directly gives the energy of
the configuration.  The lowest energy solution is the one with the
smallest value of $\alpha$, i.e., solution (e) in Figures
\ref{fig:zt_bfields} and \ref{fig:zt_UH}. Second, we see that the
solutions do not live at minima of the energy with respect to $\alpha$
but rather at inflection points.  This is because we determined the
value of $\alpha$ from a boundary condition at $r=R$ rather than by
minimizing the energy.

A system that reaches one of these zero-torque solutions will
conserve helicity since there is no outward flow of helicity through
the surface.  However, it is not clear how long the system will
survive in this state.  Since there are neighboring solutions (with
other values of $\alpha$) which have smaller values of the energy, the
system is likely to evolve away from the zero-torque solution.  Once
it does this, the stress at the surface will no longer vanish and
helicity will cease to be conserved.  Whether or not this sequence of
events can happen depends on the nature of the dynamical constraints
that the fluid must satisfy, a topic that is beyond the scope of this
paper.

\subsection{Explicit Expressions for the Magnetic Field} \label{EEftMF}
In this subsection we provide for convenience explicit expressions for
the magnetic field geometries of a few low-order multipole solutions,
including those presented in Figures \ref{fig:bfields}a-d and
\ref{fig:zt_bfields}e-f.

\begin{table}
\begin{center}
\caption{Approximate values of $\alpha R$ associated with various multipole
  solutions for the minimum energy (ME) and zero-torque (ZT)
  solutions, described by equations (\ref{eq:explicitBs}) and
  (\ref{eq:zt_explicitBs}).  Values associated with images shown are
  labeled by the corresponding figure number in the Notes column.} \label{tab:alphas}
\begin{tabular}{ccccc}
\hline
&&&&\\
Model Type & $n$ & $l$ & $\alpha R$ & Notes\\
&&&&\\
\hline
&&&&\\
ME & 0 & 1 & $\pi$    & Global Energy Minimum, Fig. \ref{fig:bfields}a\\
.  & 1 & 1 & $2\pi$   & Fig. \ref{fig:bfields}b\\
.  & 2 & 1 & $3\pi$   & Fig. \ref{fig:bfields}c\\
.  & 0 & 2 & $4.4934$ & Fig. \ref{fig:bfields}d\\
.  & 1 & 2 & $7.7253$ & --\\
.  & 2 & 2 & $10.904$ & --\\
ZT & 0 & 1 & $4.4934$ & Fig. \ref{fig:zt_bfields}e\\
.  & 1 & 1 & $7.7253$ & Fig. \ref{fig:zt_bfields}f\\
.  & 2 & 1 & $10.904$ & Fig. \ref{fig:zt_bfields}g\\
.  & 0 & 2 & $5.7635$ & Fig. \ref{fig:zt_bfields}h\\
.  & 1 & 2 & $9.0950$ & --\\
.  & 2 & 2 & $12.323$ & --\\
&&&&\\
\hline
\end{tabular}
\end{center}
\end{table}

A natural normalization of the expressions obtained in sections
\ref{sec:gs} is in terms of the standard multipole
field components of the surface magnetic field, $B_l$.  In this case
the exterior and interior field of a single multipole component is
given by
\begin{equation}
\begin{aligned}
&\B_{r\ge R} =  B_l\left[
\left(\frac{R}{r}\right)^{l+2} \bY_l
-
\frac{1}{l+1}\left(\frac{R}{r}\right)^{l+2} \bPsi_l
\right]\,,\\
&\B_{r<R} = B_l\bigg\{
\frac{R j_l(\alpha r)}{r j_l(\alpha R)} \bY_l\\
&\,\quad\qquad\qquad\qquad+
\frac{\alpha R r j_{l-1}(\alpha r) - l R j_l(\alpha r)}{l(l+1) r j_l(\alpha R)} \bPsi_l\\
&\quad\qquad\qquad\qquad\qquad\qquad\qquad-
\frac{\alpha R}{l(l+1)} \frac{j_l(\alpha r)}{j_l(\alpha R)} \bPhi_l
\bigg\}\,.
\end{aligned}
\label{eq:explicitBs}
\end{equation}
The field may then be explicitly constructed given a value of $\alpha
R$, and the explicit forms of the vector spherical harmonics and
spherical Bessel functions given in Tables \ref{tab:vspharms} and
\ref{tab:sphbessel}, respectively.  The values of $\alpha R$
corresponding to the first few minimum energy (ME) multipole solutions
are given in Table \ref{tab:alphas}.

However, the normalization in equation (\ref{eq:explicitBs}) is
divergent when the radial component of the surface field vanishes, and
thus for the zero-torque solutions described in section
\ref{sec:zerotorque} and shown in Figures \ref{fig:zt_bfields} \&
\ref{fig:zt_UH}.  For these a better choice of normalization is in
terms of the contribution of a given multipole to the total magnetic
helicity, which simplifies to
\begin{equation}
H^{\rm zs}_l
=
\frac{B_l^2 R^4}{8\pi l(l+1)} \alpha R \frac{j^2_{l-1}(\alpha R)}{j^2_l(\alpha R)}\,,
\end{equation}
and also diverges for the zero-torque configurations [$j_l(\alpha R)=0$] 
when the strength of the radial component of the surface field ($B_l$)
is held fixed. Explicitly, in terms of the contribution to the
helicity, the solutions with vanishing surface stresses are given by
\begin{equation}
\begin{aligned}
&\B_{r\ge R} = 0\,,\\
&\B_{r<R} = 
\sum_l \sqrt{\frac{8\pi l(l+1)}{\alpha R^5}  H^{\rm zs}_l}
\bigg\{
\frac{R j_l(\alpha r)}{r j_{l-1}(\alpha R)} \bY_l\\
&\,\quad\qquad\qquad\qquad+
\frac{\alpha R r j_{l-1}(\alpha r) - l R j_l(\alpha r)}{l(l+1) r j_{l-1}(\alpha R)} \bPsi_l\\
&\quad\qquad\qquad\qquad\qquad\qquad\qquad-
\frac{\alpha R}{l(l+1)} \frac{j_l(\alpha r)}{j_{l-1}(\alpha R)} \bPhi_l
\bigg\}\,.
\end{aligned}
\label{eq:zt_explicitBs}
\end{equation}
Note that since we used the vanishing surface stress limit for the
helicity, the above expression is only valid for the configurations
discussed in section \ref{sec:zerotorque}.  A more general expression
can be normalized in terms of the helicity
[eq. (\ref{eq:helicity_ff})], however this is correspondingly more
complicated.  The values of $\alpha R$ associated with the first few
multipoles of the zero-torque (ZT) configurations are given in Table
\ref{tab:alphas}.

\subsection{Star With a Rigid Crust} \label{sec:crust}

We now consider a star with a rigid crust that is able to support
unbalanced magnetic surface stresses.  The particular astrophysical
system we have in mind is a magnetized neutron star.  A rigid crust
provides a firm anchor for any field lines that penetrate it.  As a
result, kinks in the magnetic field at the surface can survive
indefinitely and there is no tendency for helicity to flow out to
infinity.  An additional effect of the crust is that any radial field
component that is present at the stellar surface when the crust first
forms will survive unchanged for a long time.  This permanent surface
field serves as a boundary condition at $r=R$.

We are then led to pose the following new problem: Given a fixed
structure of the radial magnetic field at the stellar surface (i.e.,
values of the coefficients $a_l$) and a fixed total helicity of the
system (the value of $H$), what are the equilibrium configurations of
the magnetic field?

The magnetic field distribution exterior to the star is uniquely
determined by the coefficients $a_l$.  The solution in the interior of
the star must be force-free and must
therefore take the form (\ref{eq:ffB}), with the coefficients $d_l$
given by the matching condition (\ref{eq:match}).  Therefore, the only
free parameter in the solution is $\alpha$, and its value must be
chosen such that the system has the required helicity $H$.  The
problem is thus quite similar to that posed in \S\ref{sec:zerotorque}.
The only difference is that, instead of having a vanishing field at
the stellar surface, we now have a prescribed non-vanishing
distribution of field strength at the surface.

It is useful to consider a specific numerical example.  In the
following, we have chosen the $a_l$ to scale as $l^{-5/6}/l(l+1)$, and
thus $B_l^2\propto l^{-5/3}$, to
crudely model the effect of a Kolmogorov turbulent cascade.  Given a
set of $a_l$ and a selected value of $\alpha$, equation
(\ref{eq:helicity_ff}) allows us to calculate the helicity of each
multipole and equation (\ref{eq:energy_ff}) gives the corresponding
energy.  Moreover, since the vector spherical harmonics are mutually
orthogonal, the total helicity and energy of the system are obtained
by simply summing the contributions from the individual multipoles.

\begin{figure}
\begin{center}
\includegraphics[width=\columnwidth]{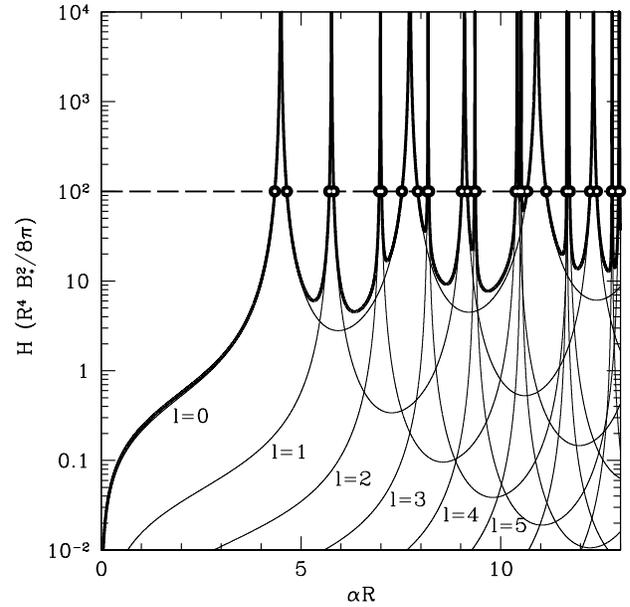}
\end{center}
\caption{The helicity as a function of $\alpha R$ in units of the
  $B_*^2 R^4/8\pi$ where $B_*$ is the normalization of the surface
  dipole field.  The thin solid lines are the helicities of the
  individual modes, normalized by the Kolmogorov scaling described in
  the text.  The thick solid line is the total helicity associated
  with the Kolmogorov spectrum.  Also shown is a constant helicity of
  100 by the long-dashed line, which crosses the helicity curve at a
  number of points.  Note that the spikes in the helicity are due to
  the fact that we have fixed the magnitude of the exterior field,
  which implies that the interior field strength diverges at the
  zero-torque configurations.}
\label{fig:rc_H}
\end{figure}

\begin{figure}
\begin{center}
\includegraphics[width=\columnwidth]{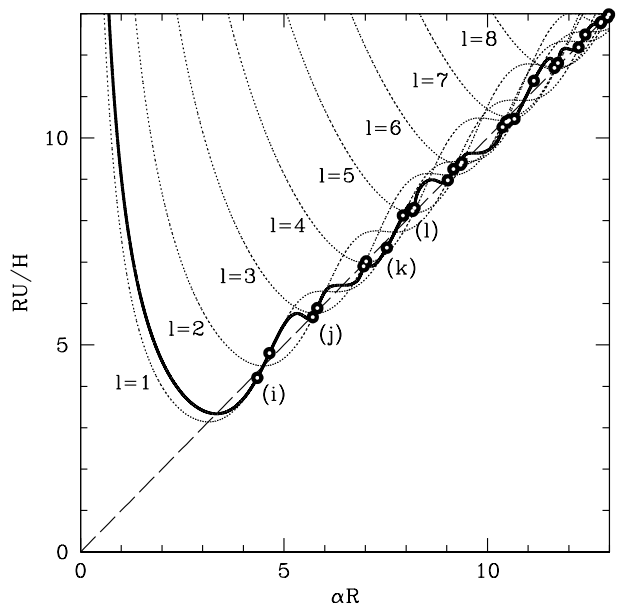}
\end{center}
\caption{$U/H$ as a function of $\alpha R$ for the fixed Kolmogorov
  surface field spectrum described in the text.  The points are states
  with $H=100$ as shown in Figure \ref{fig:rc_H}, and those that are
  explicitly shown in Figure \ref{fig:rc_bfields} are labeled (i),
  (j), (k) and (l).  Note that these are not necessarily energy
  minima, which is a direct result of fixing the multipolar structure
  at the stellar crust.}
\label{fig:rc_UH}
\end{figure}

\begin{figure*}
\begin{center}
\begin{tabular}{cc}
\includegraphics[width=0.5\textwidth]{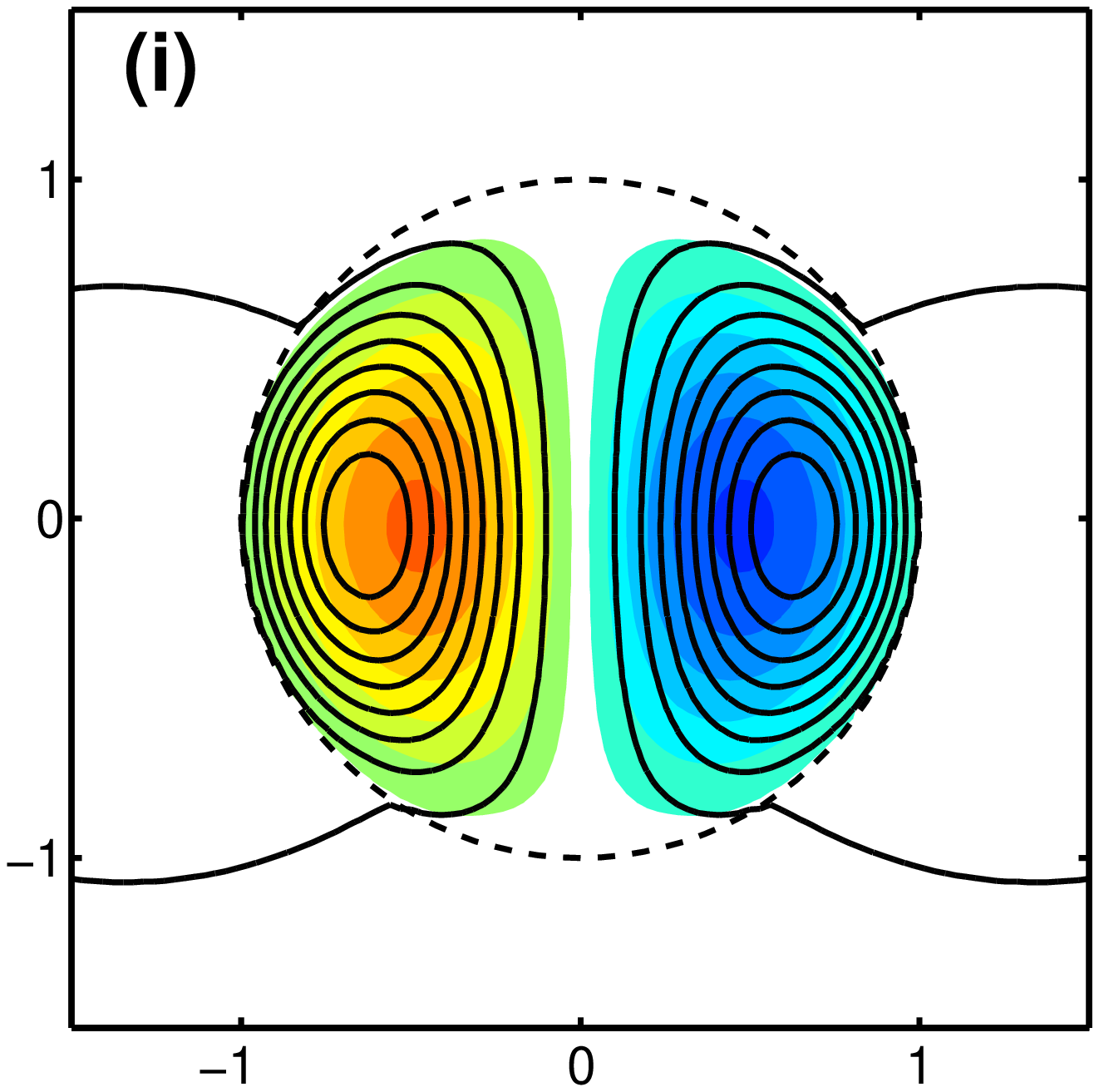}
&
\includegraphics[width=0.5\textwidth]{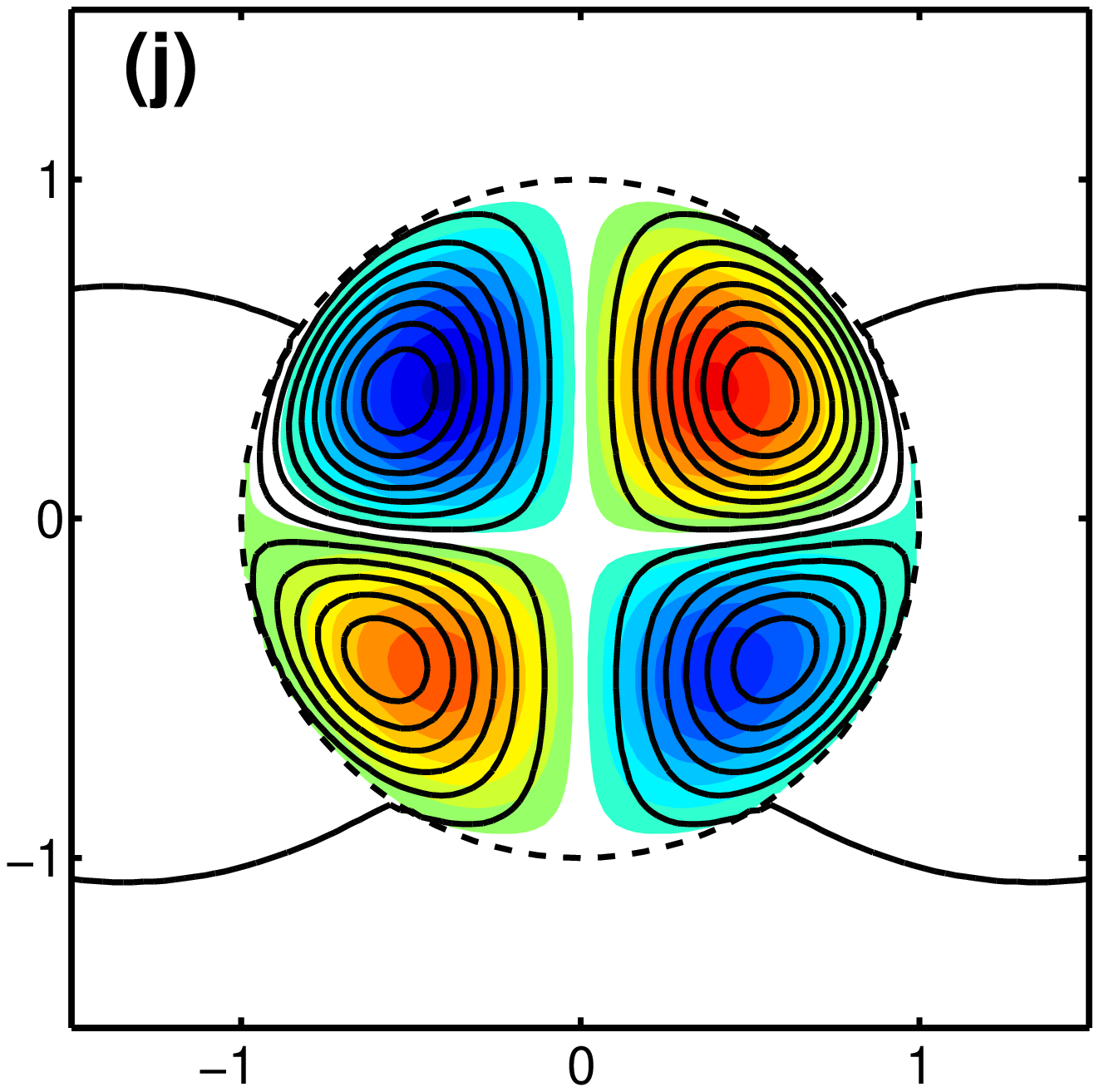}
\\
\includegraphics[width=0.5\textwidth]{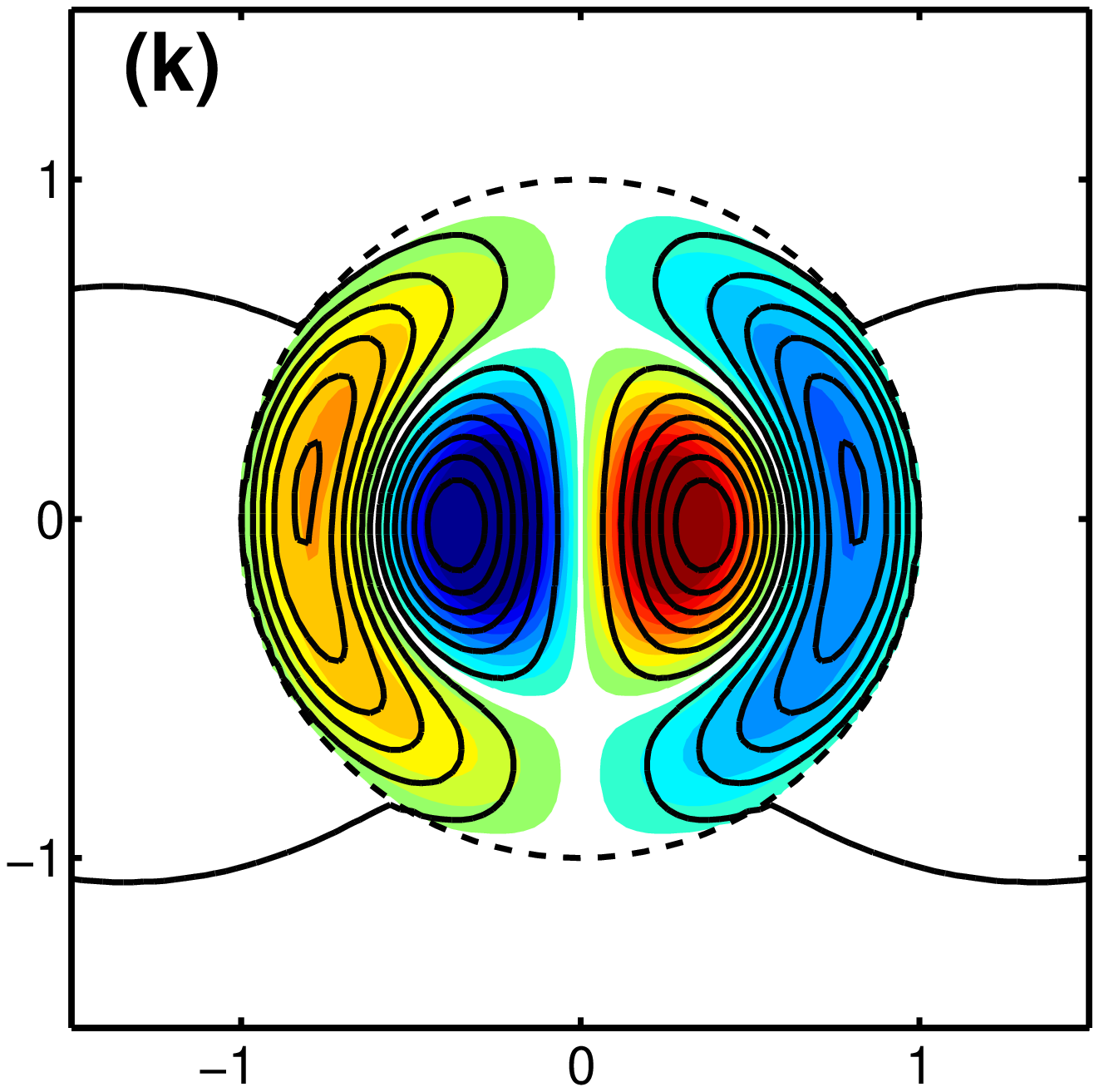}
&
\includegraphics[width=0.5\textwidth]{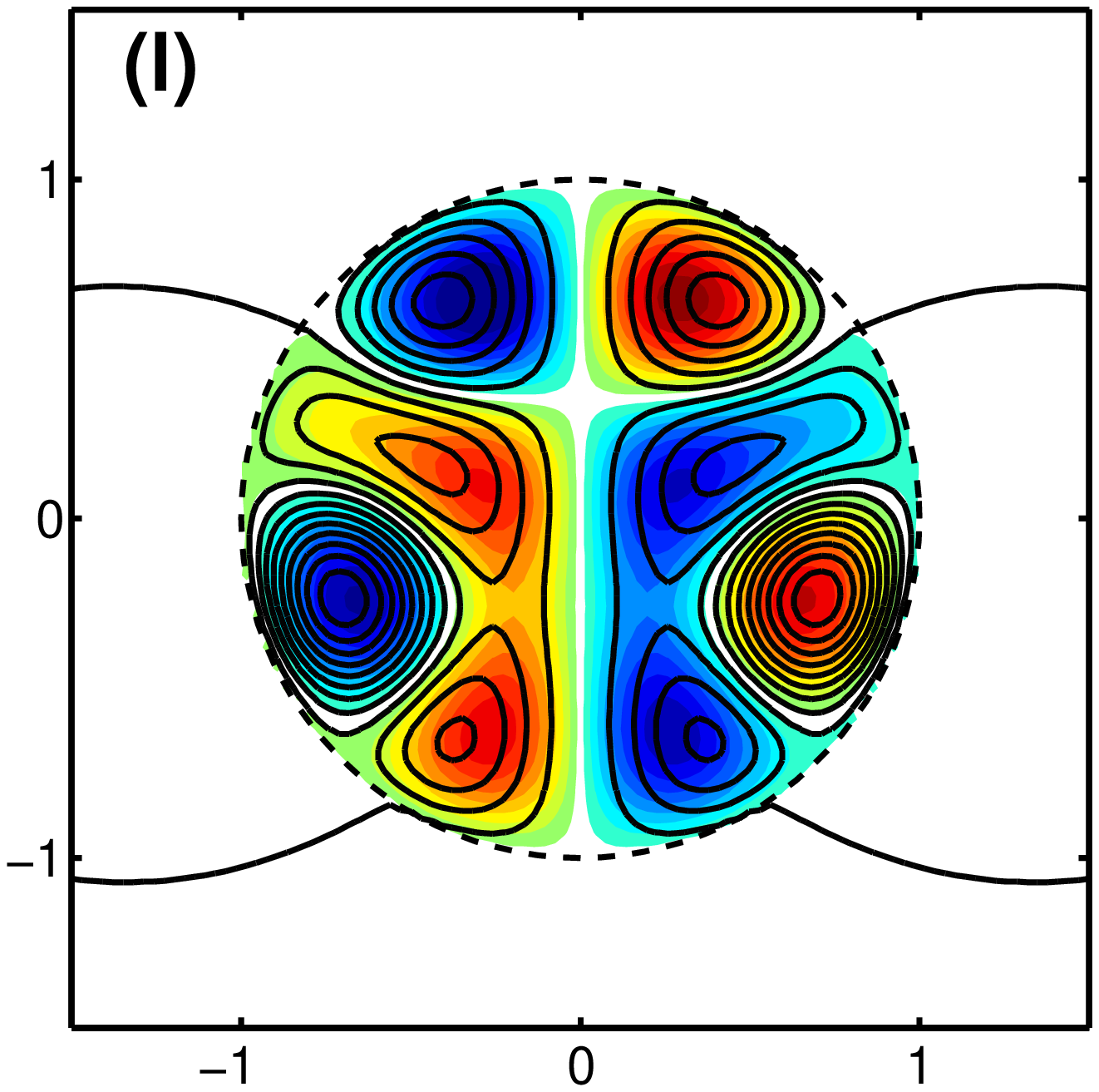}
\end{tabular}
\end{center}
\caption{Meridional slices of a particular realization of the
  Kolmogorov spectrum described in the text for the states marked in
  Figure \ref{fig:rc_UH}, each corresponding to a helicity of 100.  In
  particular, (i), (j) and (k) show the first, third and seventh
  lowest energy states with a helicity of 100, while (l) shows the
  more complicated structures that can appear in higher energy
  states.  The poloidal and toroidal field structure are shown by the
  field lines and filled-color contours, respectively.  In all plots
  the field line density is proportional to the poloidal flux
  densities and the color scheme is proportional to the toroidal flux
  densities.  Both are normalized such that all frames have identical
  total helicities.  Note the similarities with Figure
  \ref{fig:zt_bfields}.  Note also that all the panels have identical
  exterior field distributions.}
\label{fig:rc_bfields}
\end{figure*}

Figure \ref{fig:rc_H} shows the variation of total helicity (in appropriate units)
as a function of $\alpha R$ for the numerical example described above.
The many spikes in the curve correspond to the various zeros of
$j_l(\alpha R)$; as equation (\ref{eq:helicity_ff}) shows, the
helicity diverges at these zeros.  If the system has a certain
conserved helicity $H$, which we have chosen to be 100 in the
particular example shown in Figure \ref{fig:rc_H}, there will in general be
multiple solutions for $\alpha$ which satisfy this constraint.  Each
of these is a valid solution to the problem.  The solutions come in
pairs, each bracketing a particular zero of a particular spherical
Bessel function.

Figure \ref{fig:rc_UH} shows the energies of these solutions as given by equation
(\ref{eq:energy_ff}).  We see that the solutions are close to, and
bracket, the zero-torque solutions shown in Figure \ref{fig:zt_UH}.  There is thus a
close connection between the rigid-crust solutions and the
zero-torque solutions.  Finally, Figure \ref{fig:rc_bfields} shows meridional
slices of the magnetic field corresponding to the four solutions
identified as (i), (j), (k) and (l) in Figure \ref{fig:rc_UH}.  The family resemblance of
these solutions to those shown in Figure \ref{fig:zt_bfields}, and
indeed also Figure \ref{fig:bfields}, is obvious.


The strong similarity between the rigid-crust and zero-torque
solutions is easy to understand.  We have selected a somewhat large
value of $H$ for the former, which means that the solutions must have
a significantly larger field strength in the interior of the star
compared to the surface, as is evident in Figure \ref{fig:rc_bfields}.
In this situation, holding the surface field fixed at a small value
(rigid-crust solutions) is nearly the same as setting the surface
field to zero (zero-torque solutions).

\subsection{Long Term Field Evolution} \label{LTFE}
In the presence of a non-zero resistivity, the magnetic field will
necessarily decay.  Since, in practice, the resistive decay timescales
are typically much longer than the Alfv\'en crossing time, we may
approximate the state of the magnetic field as a series of
non-equilibrium configurations, for which the general solution is
given by equations (\ref{eq:ffB}) and (\ref{eq:vacB}).  Furthermore,
we will assume that the field begins in a local minimum energy
configuration, where the large-scale structure prevents further fast
reconnection.

As the field evolves due to resistive decay, the energy per unit
helicity, $U/H$, will change as well.  As a consequence, the value of
$\alpha$ will evolve.  This will generally result in the evolution of
the magnetic field structure, provided the resistivity is not
scale-invariant, \ie, small scale currents damp more rapidly than
large scale currents.  To see this consider Ohmic decay with a
constant resistivity, $\rho$.  In this case, the local field strength
evolves according to
\begin{equation}
\dot{\B} = -\curl\E = -\rho\curl\J
= -\frac{\rho}{4\pi}\curl\curl\B = -\frac{\alpha^2\rho}{4\pi}\B\,.
\end{equation}
Upon inserting equation (\ref{eq:ffB}) into this, we find
\begin{equation}
\dot{a}_l = - \frac{\alpha^2\rho}{4\pi} a_l
\qquad\text{and}\qquad
\dot{\alpha} = 0\,,
\end{equation}
\ie, the magnetic field strength (proportional to $a_l$) decreases,
but the structure (described by $\alpha$) remains unchanged.

However, if the poloidal currents decay more rapidly than the
toroidal currents, \eg, due to enhanced dissipation of the surface
currents or because dissipation is non-linear, $\alpha$ will be time
dependent.  This is clear from equation (\ref{eq:ffB}), in which
$B^\phi / B^r \propto \alpha$.  Thus if poloidal currents (and hence
toroidal magnetic fields) decay more rapidly than toroidal currents
(and hence poloidal magnetic fields), $\dot{\alpha}<0$ and the
magnetic field configuration will move to the left in Figures
\ref{fig:UH}, 4 and 6.  When the system evolves out of a local
equilibrium state a substantial amount of magnetic energy is
necessarily liberated.  The natural timescale over which this energy
is released is the Alf\'en crossing time, though we do not address
the details of the instability here.


Finally, note that in the presence of a conducting magnetosphere with
an infinitely conductive surface a fluid star would appear as a
uniform medium.  This is despite the fact that the magnetosphere has a
much lower density than the stellar interior, and thus is always
described by force-free dynamics.  This disparity results in very
different dynamical times for the two regions.  Nevertheless, since
currents can freely flow between the stellar interior and the
magnetosphere, the minimum energy equilibrium configuration will be
described by that for a uniform medium, which has no stable
equilibrium magnetic field configuration for any value of magnetic
helicity.  That is, in this case $\alpha = U/H$ can take any value,
including zero.  Thus such a field is unconditional unstable.  Whether
or not this occurs in the context of a stellar field depends upon the
conductivity of the surface, the exterior, and the ability of the
crust to fix the magnetic field.

\section{Discussion and Application to Neutron Stars}
\label{A}

\subsection{General Discussion}

Although we have described three different kinds of solutions ---
minimum-energy solutions (\S\ref{sec:minenergy}), zero-torque
solutions (\S\ref{sec:zerotorque}) and rigid-crust solutions
(\S\ref{sec:crust}) --- all have certain features in common.

For each class of solutions, the lowest energy state (``ground
state''), subject to a fixed total helicity, is dominated by a
dipolar field configuration.  Nevertheless, within the context of a
magnetic relaxation model, the presence of higher order multipoles is
not unexpected.  After the cessation of a dynamo in the early history
of a star, the resulting magnetic field will be chaotic and highly
tangled over a wide range of scales.  The field will also be violently
unstable, decaying rapidly towards a local minimum energy
configuration.  However, as described in \S\ref{LTFE}, this will only
proceed as long as the rearrangement is driven by small scale
reconnection.  Large scale rearrangements, such as converting
low-order multipoles into dipoles, will occur over the much longer
resistive timescale.  As a consequence, we generally expect the
resulting field to be dominated by a dipole component with significant
quadrupolar and octupolar components.  This is consistent with the
outcome of numerical simulations \citep{Brai-Spru:04}.


A novel feature of our work is the identification of higher-energy
quasi-equilibrium states.  At late times, small-scale current decay
will drive the system to evolve towards smaller $\alpha$, necessarily
resulting in periods of violent instability when the system hops from
one quasi-equilibrium state to the next.  These will result in large
scale reorganizations of the internal and external stellar magnetic
fields over Alfv\'enic timescales.  As a consequence, these events
could power active periods in which a substantial fraction of the
stellar magnetic energy is released.

Since the quasi-equilibria of the dipole state are located at zeros of
low-order spherical Bessel functions, the magnetic energy release in
these events should be roughly quantized (though the system may jump
over many local equilibria in a single event), releasing an energy of
order $k \pi H/R$ for some integer $k$.  The observed energy release
will depend strongly upon the emission mechanism and will likely not
appear strictly quantized.  Nevertheless, this model implies a lower
limit upon the energy that can be released during an outburst, simply
given by the energy of the global minimum energy configuration.

The minimum-energy solutions described in \S\ref{sec:minenergy} and
Figures 1, 2, have a toroidal kink in the magnetic field at the
stellar surface, in apparent contrast to numerical simulations
\citep[\cf][]{Yosh-Yosh-Erig:06,Brai-Spru:04}.  The toroidal field
discontinuity will lead to a non-radial surface stress, which cannot
be balanced by a fluid star.  Thus, these solutions are probably not
of practical interest.

The zero-torque solutions described in \S\ref{sec:zerotorque} and
Figures \ref{fig:zt_bfields}, \ref{fig:zt_UH}, are more promising.  They are perfectly force-free
solutions with no unbalanced stresses.  However, they have some
deficiencies.  First, they have no magnetic field outside the star.
Therefore, they are not very useful for understanding the observed
exterior field of systems such as magnetic A stars and white dwarfs.
Second, these solutions are not minimum energy states of the
force-free problem.  Therefore, a star in one of these states could,
in principle, lower its energy by evolving to a nearby configuration
with a lower value of $\alpha$.  The neighboring lower-energy state
will not be stress-free at the surface and the system may thus develop
a dynamical instability.  Whether or not this will happen in a real
system depends on dynamical constraints which we cannot address.
Modulo this caveat, the zero-torque solutions are of possible interest
for understanding the numerical work of \citet{Brai-Spru:04}.

All the solutions we have described in this paper have a constant
$\alpha$ in the stellar interior.  This is a natural consequence of
the variational approach we have taken, which is based on the
principle of conservation of total helicity \citep{Tayl:74}, but it is
not necessarily appropriate for a magnetized star.  We could, in
principle, consider solutions in which a part of the stellar interior
has a non-zero value of $\alpha$ and contains all the toroidal field
and helicity, while the rest of the star has $\alpha=0$ and purely
poloidal structure.  Such solutions are more difficut to construct,
but may be able to produce stable configurations with no stress at the
surface.  An even more extreme scenario is to consider a separately
conserved microscopic helicity for each magnetic flux tube (see
Appendix A).  These topics are beyond the scope of this paper.

\subsection{Neutron Stars}\label{sec:neutronstar}
During their birth, neutron stars are expected to have convectively
driven dynamos and to develop extremely strong magnetic fields, as
high as $10^{15-16}\,\G$.  However, shortly afterward, and prior to
the development of a solid crust, the dynamo action ceases and the
nascent neutron star magnetic field evolves for roughly $100\,\s$.
During this time magnetic relaxation will play a central role in
determining the strength and structure of the magnetic field.  Once
the crust forms, the magnetic field at the surface of the star is
frozen in and the magnetic helicity of the system is fixed.  Further
evolution of the system is likely governed by total helicity
conservation, coupled with the requirement that the radial field at
the stellar surface is fixed.  These are precisely the conditions we
have assumed in deriving the rigid-crust solutions discussed in
\S\ref{sec:crust} and Figures 5--7.

Generally, we expect a dominant dipole field structure, with
non-negligible quadrupole and octupole components.  Since the magnetic
helicity is a {\em scalar} quantity, we naturally expect a non-aligned
magnetic field with respect to the rotation axis.  For a highly
tangled initial field (with considerable power on small-scales), we
would expect the orientation of the birth fields of neutron stars to
be uniformly distributed relative to the spin axis.  This provides a
natural way in which to produce the off-axis fields required for
understanding radio and X-ray pulsars.

For a system with non-zero total helicity, since any magnetic field at
the stellar surface will generally have an associated toroidal stress,
we speculate that the fluid neutron star before the formation of the
crust will evolve through states in which the surface field is
generally weak relative to the field in the interior.  This condition
will be frozen in once the crust forms.  Thus, the frozen-in helicity
will be large compared to the characteristic helicity we might expect
based on the magnitude of $B_s$, the surface field strength: $H \gg
(B_s^2/8\pi)R^4$.  Similarly, the magnetic energy will also be large:
$U \gg (B_s^2/8\pi)R^3$.  These are the conditions we assumed for the
examples shown in Figures 5--7.

Once the crust forms, the system will settle down into the nearest
available solution, which would presumably correspond to a relatively
large value of $\alpha$.  Subsequently, we imagine that the magnetic
field will evolve via sudden jumps from one solution to the next, with
a steady reduction in the magnitude of $\alpha$.  Each jump will cause
a decrease in the total magnetic energy and a corresponding increase
in the thermal energy of the star.  The leaking of this thermal energy
out of the star will power thermal X-ray emission from the surface.
This may explain the quasi-steady X-ray emission seen in anomalous
X-ray pulsars (AXPs) and other magnetars.

It is interesting to note that the source of this thermal energy is by
its very nature non-steady.  Bursts of energy are released each time
the interior magnetic field switches from one state to the next.
Although the flux escaping from the surface will be smoothed on the
thermal diffusion time, some variability is expected and is indeed
seen in magnetars.

Furthermore, as we see from Figures 5 and 6, neighboring solutions are
bunched more closely together for larger values of $\alpha$ and are
farther apart at lower $\alpha$.  Since the magnetic energy is to a
good approximation proportional to $\alpha$ (see Fig. 6), we expect
frequent small bursts of energy early in the life of the neutron star,
when $\alpha$ is large, and rarer but more powerful bursts later on.
Thus, there should be less variability in younger systems and larger
fluctuations in older systems, with the most dramatic variations
occurring in the oldest magnetars.

The above discussion was concerned with magnetic field evolution in
the interior of the star with a fixed surface field.  However, since
the star has a much sronger field in its interior compared to the
surface, we expect the crust occasionally to crack under the pressure
and to release some of the enclosed energy and helicity in a sudden
cataclysmic event.  These events dump a lot of energy near the surface
of the star and provide a natural mechanism for powering bursts
associated with soft-gamma repeaters (SGRs).  With typical energies
ranging from $10^{36-46}\,\erg$, SGR bursts require interior magnetic
fields on the order of $10^{9-14}\,\G$, which are considerably less
than the equipartition values that are possible as a result of
magnetar formation \citep{Thom-Dunc:93,Spru:02}.  For magnetars, the
Alfv\'en crossing time is on the order of a second, corresponding to
maximum luminosities on the order of $10^{46}~{\rm erg\,s^{-1}}$.  If
this energy is thermalized in the neutron star crust over the Alfv\'en
timescale, it would result in surface temperatures on the order of
roughly an $\MeV$, similar to what is seen in SGR flares.  However,
the precise details of the emission will depend upon the details of
the emission mechanism \cite[see, \eg,][]{Thom-Lyut-Kulk:02,Lyut:06}.

As in the case of internal field rearrangements, we expect the most
violent events to occur somewhat late in the life of a magnetar, when
the interior field is dominated by lower multipoles.  Thus, the
strongest outbursts should be seen in relatively old systems.  All of
these arguments suggest that SGRs as a class might be older than AXPs.

In this picture magnetars are
born with magnetic fields in configurations that are local equilibria
at strengths orders of magnitude larger than the global minimum.  If
we assume that the minimum energy of observed SGR bursts is indicative
of the energy scale of the power source, this implies that the field
strength in this minimum energy state is roughly $10^9\,\G$.  Hence,
the SGR bursts would be the result of the decay of the internal
magnetic field, resulting finally in a stable field strength
commensurate with millisecond pulsars.  Note that this is a different
mechanism than that described in \citet{Brai-Spru:06}.




\section*{Acknowledgements}
This work was supported in part by NASA grant NNG04GL38G.  The authors
thank the referee for helpful criticism.  A.E.B. acknowledges the
support of an ITC Fellowship from Harvard College Observatory and
thanks Jon McKinney and Niayesh Afshordi for a number of useful
discussions.

\appendix
\section{Relationship to Other Variational Principles} \label{RtOVP}
We present a particular variational principle, with the somewhat
surprising result \citep[though similar to that found in ][]{Wolt:62} that
the minimum energy states of the magnetic field correspond to a
completely force-free state in the stellar interior with uniform
force-free constant, $\alpha$.  This is surprising since equilibrium
configurations that balance magnetic stresses against fluid pressure
are conspicuously absent from our approach.  Thus it is worthwhile to
discuss the particular conditions that our solution is valid under and
why this differs from those derived under the assumption of ideal MHD.

Our variational principle differs explicitly from that described in
\citealt{Fiel:86} (an interesting paper brought to our attention by
the anonymous referee).  The latter reproduces the standard MHD
relation, balancing pressure against magnetic stresses, leading the
author to argue that the result of \citet{Wolt:62} is only valid if
the pressure is uniform throughout the fluid.  This variational
principle is derived in a fashion nearly identical to that presented
in section \ref{HMEatFFC}, minimizing the total energy (fluid {\em
  and} magnetic) subject to a condition upon the helicity. However
there are two important distinctions.

The first concerns the terms in the fluid component of the energy.
\citet{Fiel:86} ignores the contributions due to
self-gravity.  As a consequence pressure gradients can only be
balanced by the Lorentz force.  This generally will result in an
unconditionally unstable system for isolated magnetized fluids.  In
contrast, a self-gravitating magnetized fluid can balance pressure
gradients against the gravitational acceleration, and thus the Lorentz
force may in principle vanish.  Therefore, when we include
self-gravity, the existence of pressure gradients is not synonymous
with the magnetic field violating force-free \citep[see, \eg,~][]{Chan-Pren:56}

The second concerns the way in which helicity conservation is
imposed.  In ideal MHD it is possible to define a conserved
microscopic magnetic helicity associated with each magnetic flux
tube.  Therefore, we may in principle seek to impose conservation of not
only the macroscopic, volume integrated helicity, but the helicity
associated with each field loop separately, i.e. take $\alpha$ not to
be a constant, but to be a function of position.  This is precisely
the constraint imposed in \citet{Fiel:86}.

However, it is far from clear that in the presence of large-scale
dissipation that this is the correct prescription.  Indeed, the
conjecture by \citet{Tayl:74}, apparently supported by spheromak
experiments, is that this be replaced with a volume-integrated
constraint, of the form we have employed.  The reason given by
\citet{Tayl:74} for this is that the microscopic helicity associated
with each flux tube is {\em not} conserved during reconnection events.
Indeed, during reconnection, flux tubes with different helicities may
be combined, changing the total helicity of the new loop (the precise
manner in which this occurs is not clearly defined since the
evolutionary equation for the flux-tube helicity depends upon the way
in which helicity proceeds).  However, reconnection will not change the
macroscopic topology of the magnetic field, and therefore the total
helicity will be preserved.

An alternative way to justify this is to note that if the
magnetic field rapidly reconnects, it is impossible to define unique
magnetic field loops.  That is, after sufficient time each magnetic
field loop will have been connected with every other remaining loop at
some point in the reconnection history.  As a consequence, the
force-free constant (i.e., $\alpha$) associated with each loop will
necessarily be the same, since this is constant along magnetic field
lines.  This will remain to be the case unless the helicity along some
subset of field lines can preferentially decay (which is the case in
the absence of a rigid crust).

These differences explain the absence of solutions which balance
magnetic stresses and pressure gradients in our formalism.  Such
solutions require additional constraints, such as flux freezing in
ideal MHD, which inhibit the movement of magnetic field lines through
the fluid.  By considering the high-dissipation/rapid-reconnection
limit via our volume constraint upon the total magnetic helicity we have
necessarily ignored such constraints.  That is, in the absence of such
additional constraints {\em there are no stable configurations which balance
  magnetic stresses against pressure gradients, since the fluid and
  magnetic field will move through each other until such gradients are
  erased}.  (This is identical to equilibrium configuration of a
two-component ideal gas, in which both species separately satisfy the
equilibrium condition.)  In the rapidly reconnecting limit, the
magnetic field can globally reconfigure on short timescales, and thus
we are justified in treating the magnetic field and fluid as
non-interacting in this fashion.  Finally, because we are restricting our
attention to gravitationally bound objects, non-vanishing pressure
gradients may be balanced against self-gravity, and therefore do {\em
  not} require violations of the force-free condition to maintain
equilibrium.

\section{Force-Free Fields in Uniform Media} \label{FFFiUM}

When the medium is uniformly infinitely conductive (i.e., the currents
required by the resulting magnetic field satisfy any physical
restrictions) the quantity to be minimized is
\begin{equation}
S = \int_V \frac{1}{8\pi} \left( \B\cdot\B - \alpha \A\cdot\B \right) \dV\,,
\end{equation}
where the constraint upon the total helicity is included via the
Lagrange multiplier $\alpha$.  As originally discussed in
\citet{Wolt:58} (and thus frequently referred to as the Woltjer state)
variations with respect to $\A$ produce the familiar force-free
equation $\curl\B=\alpha\B$ for the magnetic field.  That
this produces a force-free field is not unexpected; non-force-free
perturbations to a force-free solution necessarily require work to be
done upon, and thus energy added to, the magnetic field.  For the
purpose of comparison with what follows we will derive this equation
below.

First note that
\begin{equation}
\begin{aligned}
\delta(\B\cdot\B) &= 2 \delta\A\cdot(\curl\B) - 2\div ( \B\times\delta\A )\,,\\
\delta(\A\cdot\B) &= 2 \delta\A\cdot(\curl\A) - \div ( \A\times\delta\A )\,.
\end{aligned}
\end{equation}
Therefore,
\begin{equation}
\begin{aligned}
8\pi \delta S
=&
\int_V \delta\A \cdot \left( \curl\B - \alpha\curl\A \right) \dV\\
&-
\oint_{\partial V} \left( \B\times\delta\A - \frac{\alpha}{2}
  \A\times\delta\A \right)\cdot\dS\\
=&
\int_V \delta\A \cdot \left( \curl\B - \alpha\B \right) \dV\\
&-
\oint_{\partial V} \delta\A \cdot \left[ \hn\times\left(\B - \frac{\alpha}{2}
  \A\right) \right]\dA\,,
\end{aligned}
\end{equation}
where $\hn$ is the normal to the surface defined by
$\partial V$ and $\dA$ is defined by $\dS\equiv\hn\,\dA$.  Note that
the surface integral is generally gauge dependent.  This can be cured
if $\delta\A$ is constrained to vanish at $\partial V$.  Since in what
follows we will set $\partial V$ to infinity and the energy in the
field will necessarily be finite, a gauge will necessarily exist in
which $\A$ vanishes on $\partial V$, and thus this constraint upon
$\partial V$ is justified.  Therefore, the desired magnetic field must
satisfy
\begin{equation}
\curl\B = \alpha\B\,.
\label{eq:ffe}
\end{equation}

Equation (\ref{eq:ffe}) implies that
\begin{equation}
\curl(\B-\alpha\A) = 0
\quad
\rightarrow
\quad
\alpha\A = \B + \alpha\grad\Lambda\,,
\end{equation}
and thus a natural gauge choice for non-zero $\alpha$ is the
``force-free'' gauge defined by $\Lambda=0$.  Note that this is also a
Lorentz gauge since $\div\B=0$.

Within the force-free gauge it is straightforward to show that the
helicity and minimum possible magnetic energy are proportional.  That
is,
\begin{equation}
H
=
\int_V \frac{1}{8\pi} \A\cdot\B\,\dV
=
\int_V \frac{1}{8\pi} \alpha^{-1} \B\cdot\B\,\dV = \alpha^{-1} U\,,
\end{equation}
and thus small $\alpha$ correspond to small $U$ for the same $H$.

\section{Boundary Conditions} \label {BC}
On the boundaries between type-I and type-II regions, the fields must
satisfy the usual boundary conditions:
\begin{equation}
\hn\cdot\left(\B_{\rm I} - \B_{\rm II}\right) = 0
\quad\text{and}\quad
\hn\times\left(\B_{\rm I} - \B_{\rm II}\right) = \frac{4\pi}{c}
\bmath{K}\,,
\end{equation}
where $\hn$ is the surface normal (directed from I to II) and
$\bmath{K}$ is the induced surface current density.  The proper
surface current can be found by explicitly constructing solutions and
minimizing the magnetic energy.  Note that this ignores any energy
associated with the surface currents themselves.

However, this should also emerge naturally
from the equations (\ref{eq:nmec}) and (\ref{eq:lam}).  The first is
obtained from $\div\B=0$, which is true by definition and may be
verified by taking the divergence of equation (\ref{eq:nmec}).  The
second is obtained via integrating equation (\ref{eq:nmec}) around an
infinitesimal loop crossing the surface.  Define a local Cartesian
coordinate system at the position of interest at the I-II boundary in
which $\hz \parallel \hn$ at the boundary, at which $z=0$.  Then, upon
integrating around a rectangular loop in the $x$-$z$ plane of length
$\ell$ in the $\hx$-direction and vanishing length in the
$\hz$-direction, we find
\begin{equation}
\begin{aligned}
\left(\B_{\rm I} - \B_{\rm II}\right)\cdot\ell\hx
&= \oint_C \B\cdot \dC
= \int_A \left(\curl\B\right)\cdot\dS\\
&= \int_A \left(\alpha\B - \curl\curl f\blambda\right)\cdot\dS\\
&= \oint_C \left( \alpha\A + \blambda\times\grad f - f\curl\blambda\right)\cdot\dC\\
&= \oint_C \left[ (1-f)\alpha\A - f \alpha\grad\Lambda
    + \blambda\times\grad f\right]\cdot\dC
\,.
\end{aligned}
\end{equation}

The third term in the integrand can be integrating after noting
$\grad f = \hz \delta(z)$ by definition, and thus
\begin{equation}
\oint_C \delta(z) \blambda\times\hz \cdot\dC
=
\left(\lambda\times\hz\right)\cdot\hz = 0\,.
\end{equation}
Despite the explicit presence of the vector potential, the first two
terms in the integrand are indeed gauge independent.  Explicitly, note
that since under the gauge transformation
$\A \rightarrow \A + \grad \Lambda'$ also implies that
$\Lambda \rightarrow \Lambda - \Lambda'$, and hence
\begin{equation}
\begin{aligned}
(1-f)\alpha\A - f\alpha\grad\Lambda
\rightarrow
(1-f)\alpha\A - f\alpha\grad\Lambda+\grad\Lambda'\,.
\end{aligned}
\end{equation}
Finally, noting that $\int_C \grad \Lambda' \cdot \dC = 0$ completes
the proof.

Therefore,
\begin{equation}
\left(\B_{\rm I}-\B_{\rm II}\right)\cdot\ell\hx
=
\alpha \A_{\rm I}\cdot\ell\hx + \alpha\ell\hx\cdot\grad\Lambda
\end{equation}
With an analogous expression for a loop in the $y$-$z$ plane, the
surface current is given by
\begin{equation}
\frac{4\pi}{c}\bmath{K} = \alpha \hn\times\left( \A_{\rm I} + \grad\Lambda \right)\,.
\end{equation}
If we choose the force-free gauge within regions of type-I, continuity
of the vector potential gives
\begin{equation}
\frac{4\pi}{c}\bmath{K} = \hn\times\left( \B_{\rm I} + \alpha \grad\Lambda \right)\,,
\end{equation}
or,
\begin{equation}
\hn\times\left( \B_{\rm II} - \alpha\grad\Lambda \right) = 0\,,
\end{equation}
on the boundary.  That is, as expected $\B_{\rm II}$ is a potential
field, though otherwise the surface currents are unconstrained.

\section{The Action in Non-Uniform Media} \label{TAiNUM}
In non-uniform medium, it is possible to explicitly reduce the
computation of the action to integrals over the boundaries between
conducting (type-I) and insulating (type-II) regions.  To do this,
first note that the magnetic energy is given by
\begin{equation}
\begin{aligned}
U &=
\int_{\rm I} \frac{1}{8\pi} \B\cdot\B \,\dV
+ 
\int_{\rm II} \frac{1}{8\pi}\B\cdot\grad\varphi \,\dV\\
&=
\int_{\rm I} \frac{1}{8\pi} \B\cdot\B \,\dV
-
\oint_{\partial({\rm I-II})} \frac{1}{8\pi}\varphi\B\cdot\dS\,,
\label{eq:Utot}
\end{aligned}
\end{equation}
where $\div\B=0$ was used.  Note that because the $\hn\cdot\B$ is
continuous across the boundary, this may be evaluated using only
type-I quantities.

If $\Lambda_{\rm L}$ is defined by
$\A = \A_{\rm L} + \grad \Lambda_{\rm L}$ where $\A_{\rm L}$ is in the
Lorentz gauge (i.e., $\div \A_{\rm L} = 0$), the magnetic helicity is
given by
\begin{equation}
\begin{aligned}
H &=
\int_{\rm I} \frac{1}{8\pi} \A\cdot\B \,\dV
+
\int_{\rm II} \frac{1}{8\pi} \A\cdot\grad\varphi \,\dV\\
&=
\alpha^{-1}\int_{\rm I} \frac{1}{8\pi} \B\cdot\B \,\dV
-
\oint_{\partial({\rm I-II})} \frac{1}{8\pi} \varphi\A\cdot\dS\\
&\qquad-
\int_{\rm II} \frac{1}{8\pi} \varphi \div\A \,\dV\\
&=
\alpha^{-1}\int_{\rm I} \frac{1}{8\pi} \B\cdot\B \,\dV
-
\alpha^{-1}\oint_{\partial({\rm I-II})}\frac{1}{8\pi} \varphi\B\cdot\dS\\
&\qquad-
\int_{\rm II} \frac{1}{8\pi} \varphi \nabla^2 \Lambda_{\rm L} \,\dV\\
&=
\alpha^{-1} U
+
\oint_{\partial({\rm I-II})} \frac{1}{8\pi} \varphi \grad\Lambda_{\rm L}\cdot\dS
+
\int_{\rm II} \frac{1}{8\pi} \B\cdot\grad\Lambda_{\rm L}\\
&=
\alpha^{-1} U
-
\oint_{\partial({\rm I-II})} \frac{1}{8\pi}
\left( \Lambda_{\rm L} \grad\varphi - \varphi \grad\Lambda_{\rm L} \right)\cdot\dS\,,
\end{aligned}
\end{equation}
where $\div\B=\div\A_{\rm L}=0$ and the continuity of $\A$ and
$\B\cdot\hn$ at the boundary were used (i.e., on the boundary
$\hn\cdot\A_{\rm II}=\hn\cdot\A_{\rm I}
=\alpha^{-1}\hn\cdot\B_{\rm I}=\alpha^{-1}\hn\cdot\B_{\rm II}$).
Therefore, the action may be written as
\begin{equation}
S
=
U -\alpha H
=
\oint_{\partial({\rm I-II})} \frac{\alpha}{8\pi}
\left( \Lambda_{\rm L} \grad\varphi - \varphi \grad\Lambda_{\rm L} \right)\cdot\dS\,,
\end{equation}
where $\varphi$ and $\Lambda_{\rm L}$ are implicit functions of $\alpha$
through equation (\ref{eq:nmec}).  Note that when $\Lambda_{\rm L}$
vanishes, i.e., the force-free gauge is a Lorentz gauge everywhere,
$U=\alpha H$ as found for the Woltjer state.

\section{Vector Spherical Harmonics} \label{VSH}
\begin{table*}
\begin{center}
\caption{Explicit expressions for the azimuthally symmetric
  $l=0$, $1$, $2$ and $3$ vector spherical harmonics.} \label{tab:vspharms}
\begin{tabular}{ccccc}
&&&&\\
\hline
&&&&\\
$l$ & $m$ & $\bY_{lm}$ & $\bPsi_{lm}$ & $\bPhi_{lm}$\\
&&&&\\
\hline
&&&&\\
0 & 0 & $\ds \sqrt{\frac{1}{4\pi}} \er$ & 0 & 0\\
1 & 0 & $\ds \sqrt{\frac{3}{4\pi}} \cos\theta \er$ & $\ds -\sqrt{\frac{3}{4\pi}} \sin\theta \eth$ & $\ds -\sqrt{\frac{3}{4\pi}} \sin\theta \eph$\\
2 & 0 & $\ds \sqrt{\frac{5}{16\pi}} \left(3\cos^2\theta-1\right) \er$ & $\ds -\sqrt{\frac{45}{4\pi}} \cos\theta \sin\theta \eth$ &  $\ds -\sqrt{\frac{45}{4\pi}} \cos\theta \sin\theta \eph$\\
3 & 0 & $\ds \sqrt{\frac{7}{16\pi}} \left(5\cos^3\theta-3\cos\theta\right) \er$ & $\ds -\sqrt{\frac{63}{16\pi}} \left(5\cos^2\theta\sin\theta-\sin\theta\right) \eth$ & $\ds -\sqrt{\frac{63}{16\pi}} \left(5\cos^2\theta\sin\theta-\sin\theta\right) \eth$\\
&&&&\\
\hline
\end{tabular}
\end{center}
\end{table*}
Vector spherical harmonics are used extensively,
and thus some of their properties are summarized below
\citep[see, \eg, ][ for more detail]{Barr-Este-Gira:85}.  While it
is possible to expand the angular dependence of a vector in many ways,
simplifications of the Helmholtz equation can be obtained with the
following choices:
\begin{equation}
\bY_{lm} \equiv \er Y_{lm}\,,\quad
\bPsi_{lm} \equiv r \bnabla Y_{lm}\,,\quad
\bPhi_{lm} \equiv \bmath{r}\times\bnabla Y_{lm}\,,
\end{equation}
where the $Y_{lm}$ are the standard scalar spherical harmonics (see
Table \ref{tab:vspharms} for explicit expressions for low $l$).  These
have the following useful properties:
\begin{equation}
\begin{gathered}
\div\left(F(r)\bPhi_{lm}\right) = \,0\,,\\
\curl\left(F(r)\bY_{lm}\right) = - \frac{F(r)}{r} \bPhi_{lm}\,,\\
\curl\left(F(r)\bPsi_{lm}\right) = \left(\frac{1}{r}
\partial_r r F(r)\right)\bPhi_{lm}\\
\begin{aligned}
\curl\left(F(r)\bPhi_{lm}\right) =&
-\left(\frac{l(l+1)}{r} F(r)\right) \bY_{lm}\\
&-\left(\frac{1}{r} \partial_r r F(r)\right)\bPsi_{lm}\,.
\end{aligned}
\end{gathered}
\label{eq:vsh_props}
\end{equation}
The are also orthogonal on the unit sphere:
\begin{equation}
\begin{gathered}
\int \d\Omega \bY_{lm}\cdot\bY^*_{l'm'} = \delta_{ll'}\delta_{mm'}\,,\\
\int \d\Omega \bPsi_{lm}\cdot\bPsi^*_{l'm'} = l(l+1)\delta_{ll'}\delta_{mm'}\,,\\
\int \d\Omega \bPhi_{lm}\cdot\bPhi^*_{l'm'} = l(l+1)\delta_{ll'}\delta_{mm'}\,,\\
\int \d\Omega \bY_{lm}\cdot\bPsi^*_{l'm'} =
\int \d\Omega \bY_{lm}\cdot\bPhi^*_{l'm'}\\
=
\int \d\Omega \bPsi_{lm}\cdot\bPhi^*_{l'm'} = 0\,.
\end{gathered}
\end{equation}

Note that equation (\ref{eq:vsh_props}) implies that
\begin{equation}
\bnabla\times\bnabla\times\left(F(r)\bPhi_{lm}\right)
=
\left( -\Delta_l F(r) \right) \bPhi_{lm}\,,
\end{equation}
where $\Delta_l$ is the 3-dimensional Laplacian associated with
the meridional harmonic $l$, i.e.,
\begin{equation}
\Delta_l = \frac{1}{r^2} \partial_r r^2 \partial_r - \frac{l(l+1)}{r^2}\,.
\tag{\ref{eq:Deltal}}
\end{equation}


\begin{table*}
\begin{center}
\caption{Explicit expressions for the spherical Bessel functions
  of orders $l=0$, $1$, $2$ and $3$ \citep{Abra-Steg:72}.} \label{tab:sphbessel}
\begin{tabular}{cccc}
\hline
&&\\
$l$ & $j_l(z)$ & $n_l(z)$\\
&&\\
\hline
&&\\
0 & $\ds \frac{1}{z}\sin z$ & $\ds -\frac{1}{z}\cos z$\\
1 & $\ds \frac{1}{z^2}\sin z-\frac{1}{z}\cos z$ & $\ds -\frac{1}{z^2}\cos z-\frac{1}{z}\sin z$\\
2 & $\ds \left(\frac{3}{z^3}-\frac{1}{z}\right)\sin z -\frac{3}{z^2}\cos z$ & $\ds -\left(\frac{3}{z^3}-\frac{1}{z}\right)\cos z - \frac{3}{z^2}\sin z$\\
3 & $\ds \left(\frac{15}{z^4}-\frac{3}{z^2}\right)\sin z -\left(\frac{15}{z^3}-\frac{1}{z}\right)\cos z$ & $\ds -\left(\frac{15}{z^4}-\frac{3}{z^2}\right)\cos z - \left(\frac{15}{z^3}-\frac{1}{z}\right)\sin z$\\
&&\\
\hline
\end{tabular}
\end{center}
\end{table*}

Note that the general solution to $\Delta_l f + \alpha^2 f = 0$ are
the spherical Bessel functions of order $l$, $j_l(\alpha r)$ and
$n_l(\alpha r)$.  The first few of these are summarized in Table
\ref{tab:sphbessel}, and more information may be found in
\citet{Abra-Steg:72}.

\bibliography{brelax.bib}
\end{document}